\documentclass[12pt]{article}%
\usepackage{amsfonts}
\usepackage{amsmath}
\usepackage{bm}
\usepackage{hyperref}
\usepackage{graphicx}
\usepackage{amssymb}
\usepackage[doublespacing]{setspace}
\usepackage{natbib}
\usepackage{geometry}
\usepackage{sectsty}
\usepackage[font=small]{caption}
\usepackage[draft]{changes}
\usepackage{subcaption}
\usepackage{booktabs}%
\providecommand{\U}[1]{\protect\rule{.1in}{.1in}}
\providecommand{\U}[1]{\protect\rule{.1in}{.1in}}
\numberwithin{equation}{section}
\newtheorem{theorem}{Theorem}

\newtheorem{algorithm}[theorem]{Algorithm}

\newtheorem{lemma}{Lemma}

\newtheorem{proposition}{Proposition}

\definechangesauthor[color=red]{Jiaqi}
\ifx\pdfoutput\relax\let\pdfoutput=\undefined\fi
\newcount\msipdfoutput
\ifx\pdfoutput\undefined\else
\ifcase\pdfoutput\else
\ifx\paperwidth\undefined\else
\ifdim\paperheight=0pt\relax\else\pdfpageheight\paperheight\fi
\ifdim\paperwidth=0pt\relax\else\pdfpagewidth\paperwidth\fi
\fi\fi\fi
\begin{document}

\title{On the numerical approximation of minimax regret rules via fictitious
play\thanks{Authors' email addresses are pxg27@psu.edu and
jiaqi.huang@psu.edu, respectively. We would like to thank Toru Kitagawa, Chuck
Manski, and Pepe Montiel for very helpful discussion and references. We thank
seminar participants in Berlin (Humboldt University), Cornell, Freiburg,
UC\ Irvine, and UCSD for helpful comments.}}
\author{$%
\begin{array}
[c]{c}%
\text{Patrik Guggenberger}\\
\text{Department of Economics}\\
\text{Pennsylvania State University}%
\end{array}
$
\and $%
\begin{array}
[c]{c}%
\text{Jiaqi Huang}\\
\text{Department of Economics}\\
\text{Pennsylvania State University}%
\end{array}
$}
\date{First version December 2024. This version \today}
\maketitle

\begin{abstract}
Finding numerical approximations to minimax regret treatment rules is of key
interest. To do so when potential outcomes are in $\{0,1\}$ we discretize the
action space of nature and apply a variant of Robinson's (1951) algorithm for
iterative solutions for finite two-person zero sum games. Our approach avoids
the need to evaluate regret of each treatment rule in each iteration. When
potential outcomes are in $[0,1]$ we apply the so-called coarsening approach.

We consider a policymaker choosing between two treatments after observing data
with unequal sample sizes per treatment and the case of testing several
innovations against the status quo.



\thispagestyle{empty}

\end{abstract}%

\setcounter{page}{1}%

\section{Introduction\label{Introduction}}

\setcounter{equation}{0}\hspace{0.25in}Consider a policymaker who has to pick
one of $T$ treatments after being informed about treatment outcomes by a
sample of size $N$ from the population. If there is not one treatment policy
$\delta$ that is best,\footnote{A treatment rule $\delta$ is a mapping from
the set of possible samples onto the space of probability distributions on the
set $\{1,...T\},$ see (\ref{set of policy rules}) for a precise definition.}
in expected outcome say, uniformly over all possible data generating processes
(DGPs) then it is not unambiguous how an optimal treatment policy should be
defined. Typically, there is not just one rule that is admissible. One could
pursue a Bayesian analysis and optimize for a given prior over the space of
all possible DGPs. One could pursue a max-min analysis that looks for a
treatment rule that maximizes over all possible rules the minimal (over all
DGPs) expected outcome. While the Bayesian route is subjective and would lead
to potentially poor finite sample performance when the prior is false, the
max-min criterion is often uninformative in the sense of all treatment
policies being max-min. A growing literature therefore focuses on finding
\emph{minimax regret} rules, see Wald (1950), Savage (1954), and Manski
(2004), that is, rules that minimize over all treatment policies the maximal
regret over all DGPs, where regret for a given policy and DGP measures the gap
between the best possible expected outcome and the expected outcome obtained
for the chosen policy. Unfortunately, there are very few examples where
minimax regret rules are analytically known and therefore in many examples of
empirical interest they cannot currently be used by
policymakers.\footnote{Exceptions include e.g. Schlag (2006), Manski (2007),
Stoye (2009, 2012), Tetenov (2012), Montiel Olea, Qiu, and Stoye (2023), Yata
(2023), Chen and Guggenberger (2024), Kitagawa, Lee, and Qiu (2024), and
additional references in these papers. The analytical finite sample results
derived in these papers apply only under very restrictive assumptions, like an
unrestricted (except for certain bounds) symmetric parameter space for the
DGPs, an equal number of observations on outcomes for every treatment, or a
normality assumption for observed outcomes in the partially identified
examples. Guggenberger, Mehta and Pavlov (2024) derive minimax policies in the
case where a policymaker is concerned with an $\alpha$-quantile rather than
expected outcome. See also Manski and Tetenov (2023). Robust Bayes methods
constitute an additional approach and have been studied in the context of
partially-identified models in Giacomini and Kitagawa (2022) and Aradillas
Fern\'{a}ndez, Montiel Olea, Qiu, Stoye, and Tinda (2024). Important
references for optimal treatment choice in an asymptotic framework include
Hirano and Porter (2009) and Kitagawa and Tetenov (2019).} Manski (2021)
concludes that \textquotedblleft The primary challenge to use of statistical
decision theory is computational.\textquotedblright\ While there have been
some contributions with regards to numerical implementation of minimax regret
rules, see e.g. Chamberlain (2000) who provides an approach via convex
minimization, the computational burden of the currently available algorithms,
at least in the types of examples considered in the paper here, is quite
overwhelming especially with the complexity of the space of DGPs, the number
of actions by the policymaker or the sample size $N$
increasing.\footnote{Meaningful progress has been made in some cases when one
is willing to search for minimax regret rules in \emph{constrained} classes of
treatment rules. See e.g. Manski (2004), Manski and Tetenov (2016, 2019,
2021), Kitagawa and Tetenov (2018), and Dominitz and Manski (2024). The latter
paper obtains a simple expression for the maximum regret of a simple intuitive
\textquotedblleft midpoint prediction\textquotedblright\ for best prediction
under square loss with missing data.}\smallskip

The objective of this paper is to provide a feasible procedure for the
numerical approximation of minimax regret rules. We first provide an algorithm
based on Robinson (1951) to approximate minimax regret rules that applies in
the case where potential outcomes $Y_{t},$ for each $t=1,...,T,$ are elements
of $\{0,1\}$ and the set of possible DGPs that nature chooses from has been
discretized. In the case where $Y_{t}\in\lbrack0,1]$ we obtain an
approximation to a minimax regret rule via the so-called coarsening method
applied to the solution from the binary case.

For the binary case, we adapt an algorithm from Robinson (1951) who considers
a zero-sum two player game in which each player has finitely many actions and
randomization over these actions is allowed for. The algorithm iteratively
updates the strategy chosen by a player. More precisely, players take turns
and each time when it is a player's turn she picks one of the finitely many
actions that is a best response to the current mixed strategy of the other
player that equals the empirical distribution of the strategies used by that
player up to that period.\ The player then updates her strategy by setting it
equal to the empirical distribution function over all best responses she
played up to that point. Robinson (1951) shows that as the number of
iterations goes to infinity the payoff that each player secures converges to
the value of the game. In the game theory literature this approach is known as
\textquotedblleft fictitious play\textquotedblright. In our scenario, the two
players are the policymaker that picks a treatment rule and nature that picks
the DGP with the payoff to the policymaker equal to negative regret. To
conform with the setup in Robinson (1951) we discretize the set of possible
DGPs that nature can choose from.\footnote{We make the discretization of
nature's action space explicit in our discussion. But note that any numerical
approach to approximate minimax regret rules via an iterative algorithm that
relies on repeatedly calculating best responses by nature through grid search
also effectively uses discretization.} We show that as the discretization gets
finer the minimax regret of the discretized game converges to the one of the
original game. In the binary case, the set of nonrandomized treatment rules
for the policymaker is automatically finite. Applied to our setting,
Robinson's (1951) result implies that the maximal regret the policymaker might
suffer approaches the minimax regret value under the sequence of treatment
rules produced by the algorithm. We consider modifications of Robinson's
(1951) approach by using updating weights proposed by Leslie and Collins
(2006) that seem to lead to faster convergence of the algorithm.\smallskip

In each step of our proposed algorithm a best response has to be calculated.
Crucially, calculating the policymaker's best response to a given strategy by
nature can be done, in many cases of interest, computationally trivially, via
Bayes' rule and comparisons of the conditional means. We do so in our main
applications. Note that the main computational bottleneck of other methods
proposed in the literature, e.g. Chamberlain's (2000) approach based on convex
optimization, is caused by the need to either evaluate the risk of all
treatment rules or the risk of a given treatment rule for all possible actions
by nature for every iteration of the algorithm. As illustrated below, in our
examples we deal with cases where the policymaker may be tasked to randomize
among $10^{10000}$ different treatment rules (or even many more). While our
approach can successfully tackle that challenge any procedure that attempts to
evaluate the risk of each treatment rule must necessarily fail due to the
computational burden.

For a given strategy of the player we calculate nature's best response via
grid search which poses the main computational hurdle. The computational
complexity of our procedure (and any other procedure we are aware of)
increases exponentially in the dimension of nature's action space. We are
unaware of any particular structure, like convexity, to nature's maximization
problem and currently use grid search to solve it.\smallskip

Importantly, at every iteration step, the algorithm produces a bound for how
far the maximal regret of the currently proposed treatment rule differs from
the actual minimax regret value under discretization of nature's action space.
This bound converges to zero as the number of iterations increases.

We also consider the case where the policymaker uses a priori information that
restricts the parameter space of nature, for example, by restricting the
possible set of means of $Y_{t}.$ We show that this can be straightforwardly
incorporated into our general framework. Also, the set of rules the
policymaker may choose from can also be restricted in our framework to
incorporate for example certain policy restrictions. The latter however may
rule out the applicability of Bayes' rule and may thus lead to a substantial
increase in the computation time.\smallskip

In the case $Y_{t}\in\lbrack0,1]$ we immediately obtain an approximate minimax
regret rule via the \emph{coarsening approach} (see e.g. Schlag (2006) and
Stoye (2009)) and an approximate minimax regret rule from the binary case. We
illustrate how to adapt the coarsening approach to various sampling designs,
arbitrary number of treatments and incorporation of a priori information on
possible DGPs by the policymaker.\smallskip

In given applications of the proposed algorithm, we suggest searching for
\emph{symmetries} in the model that allow for further reduction of computation
time. Namely, often it is possible to show that Nash equilibria of the game
between the policymaker and nature can be found in a class of treatment rules
restricted by certain symmetry conditions. This then implies that minimax
regret rules can be found as well in that restricted class.\smallskip

We study the following examples. In the first example, we consider the case of
a policymaker who needs to pick one of two treatments after having observed
$N_{t}$ observations on each treatment$,$ $t=1,2,$ where $N_{1}$ and $N_{2}$
may be different, and aims at maximizing expected outcome. First, we consider
the case where outcomes are either successes or failures, i.e. $Y_{t}%
\in\{0,1\}$. It is shown that in this case minimax regret rules can be found
in a class of treatment rules restricted by a symmetry condition that
substantially reduces the class of rules that need to be searched over. In a
nutshell, the \textquotedblleft symmetric\textquotedblright\ rules $\delta$
considered are such that the value of $\delta(w_{N})$ for a sample\emph{
}$w_{N}=(n_{1},n_{2})$\emph{ }(where $n_{t}$ denotes the number of successes
for treatment $t$ in the sample) already pins down the value of $\delta
(w_{N}^{\prime})$ for another sample\emph{ }$w_{N}^{\prime}=(n_{1}^{\prime
},n_{2}^{\prime})$ such that $n_{t}+n_{t}^{\prime}=N_{t}$ for $t=1,2.$ That
is, pinning down the values of a symmetric rule on about one half of the
arguments already determines the entire rule. We then provide an algorithm
that leverages the insights of fictitious play with the restrictions imposed
by symmetry$.$\footnote{For a very restricted subcase we are able to provide
analytical results for minimax rules, namely the case where $\min
_{t=1,2}\{N_{t}\}=0$, that is the case where one only observes data on
outcomes from one treatment.} To give some measure about the computational
complexity that is being tackled here, one can show that in the case where
$N_{1}=N_{2}=300$ there are $2^{N_{1}(N_{2}+1)/2+N_{2}/2}=2^{45300}%
\approx10^{13636.7}$ symmetric \textquotedblleft
nonrandomized\textquotedblright\ treatment rules one needs to search
over.\footnote{For symmetric rules, defined properly in (\ref{summability}),
one needs to set $\delta_{2}(n_{1},n_{2})=1/2$ when $n_{1}/N_{1}=n_{2}%
/N_{2}=1/2.$ Therefore, when we say \textquotedblleft
nonrandomized\textquotedblright\ in the context of symmetric rules we mean
$\delta_{2}(n_{1},n_{2})\in\{0,1\}$ except for the case $n_{1}/N_{1}%
=n_{2}/N_{2}=1/2.$
\par
All computations reported in the paper are done using the Gauss software on a
Dell desktop computer with an Intel(R) Core(TM) i7-14700 2.10 GHz Processor.}
By Stoye (2009, Corollary 1) it is known that the minimax regret value in that
case equals .006940. In 51 minutes of computation time with 2000 iterations,
the algorithm produces a treatment rule (a weighted average over the
$10^{13636.7}$ symmetric nonrandomized treatment rules) whose maximal regret
equals .006939! We illustrate that our algorithm can lead to substantial
improvements in terms of maximal regret after quite few iterations relative to
an empirical success rule.

The second example we study is testing innovations against a status quo. Stoye
(2009) provides analytical expressions for minimax regret rules (given as
solutions to a univariate equation) when one innovation is compared to the
status quo, but there is no known analytical solution for the case of more
than one innovation. As in the previous example, we derive certain symmetry
conditions that help reduce the computational effort and then apply the
general algorithm for that example. Additional applications appear in a
Supplementary Appendix.\smallskip

Relative to most of the extant literature on computational approaches to
approximate minimax regret rules our approach avoids the most costly step of
the algorithm that requires evaluation of risk over all treatment rules.
Namely, consider a finite set of treatment rules with $f$ elements to choose
from and denote by $p$ a discrete probability distribution over that set, i.e.
a mixed strategy. It is well known that the function $h(p)=\sup_{s\in
\mathbb{S}}R(p,s)$ is convex over the $f-1$-dimensional simplex, where $R$
denotes regret (defined in (\ref{regret}) below) and $\mathbb{S}$ denotes the
set of all possible distributions for potential outcomes. Chamberlain (2000)
suggests using tools from convex programming to find a minimum for the
function $h(p).$ However, this step is computationally costly when $f$ is
large. Instead, in our algorithm, when a best response by the policymaker
needs to be calculated for the current mixed strategy used by nature, we rely
on Bayes' rule. The computational hurdle is further reduced by dividing the
problem into a computationally feasible numerical step and the coarsening trick.

Regarding numerical approaches for approximation of minimax regret rules,
besides the early contribution by Chamberlain (2000), Masten (2023) considers
nonlinear optimization packages like KNITRO and applies them to the case of
\emph{unbalanced} samples with $T=2$ for several small sample sizes with
$\max_{t\in\{1,2\}}\{N_{t}\}\leq5.$ See Dominitz and Manski (2024) for a
comprehensive discussion of computational methods. Recently, Aradillas
Fern\'{a}ndez, Blanchet, Montiel Olea, Qiu, Stoye, and Tan (2024) consider a
particular convex optimization routine\textbf{ }called \textquotedblleft
mirror descent\textquotedblright\ to approximate $\epsilon$-minimax regret
rules. The policymaker iteratively updates her strategy using
\textquotedblleft multiplicative weights\textquotedblright\ based on the
subgradient of $h$ while nature computes a best response in each iteration to
the policymaker's current strategy. For a bounded risk function it is shown
that it takes $O(\ln f/\epsilon^{2})$ iterations to obtain an $\epsilon
$-minimax regret rule. From results in Ben-Tal, Margalit, and Nemirovski
(2001) it follows that the algorithm is optimal with respect to number of
iterations among all iterative, first-order algorithms for convex
optimization\ over the $f-1$-dimensional simplex up to the logarithmic factor
$\ln f.$ In contrast, we do not provide results on the maximal number of
iterations our algorithm requires to produce an $\epsilon$-minimax regret rule
in a broad class of problems. We are concerned about the overall computational
effort, which combines both the number of iterations and the computational
effort in each iteration, in a given application. There are examples, where
the convergence rate of fictitious play is rather slow, see e.g. Daskalakis
and Pan (2014). But in the examples we consider in the paper, we find that the
overall computational effort to obtain an $\epsilon$-minimax regret rule a is
very competitive. We provide a comparison with \textquotedblleft mirror
descent\textquotedblright\ in the context of our first example.

Daskalakis, Deckelbaum, and Kim (2024) focuses on computing the value of a
game using the so-called \textquotedblleft excessive gap
technique\textquotedblright. Using that approach requires optimization for
each iteration of the algorithm which can be challenging in the applications
that we consider given the strategy space for the policymaker is growing very
fast. An additional related contribution regarding convex optimization is
Bubeck (2015). An important reference for an algorithm to approximate a least
favorable prior distribution is Kempthorne (1987).\smallskip

Note that the general problematic one faces here is related to a testing
scenario where several competing tests are available that all control the
size. If the power curves of the tests intersect then which test to pick
becomes a non-unambiguous task. Elliott, M\"{u}ller, and Watson (2015)
consider finding tests that maximize weighted average power and in a scenario
where the null hypothesis is composite they propose algorithms that find tests
that are nearly optimal. It is shown that finding such optimal tests can be
expressed as a minimax regret problem. Therefore, the numerical algorithm
proposed in our paper might potentially be applied also in this scenario but
would require several modifications \medskip

The paper is organized as follows. Section \ref{model and approximation}
provides a general description of the numerical procedure to approximate
minimax regret rules based on fictitious play. In Subsection
\ref{discretization} the discretized model of nature's parameter space is
introduced and subsection \ref{algorithm} discusses the algorithm to
approximate minimax regret rules. Subsection \ref{coarsening subsection}
discusses how approximate minimax regret rules can be obtained in the case
where outcomes live on the unit interval via the so-called coarsening approach
of approximate minimax regret rules from the binary case. We then consider
several examples of the general theory. Section \ref{lead example} introduces
the example of treatment choice with unbalanced sample sizes. Analytical
minimax regret rules are provided for the special case where $\min
_{t=1,2}\{N_{t}\}=0.$ Subsection \ref{symmetry} shows that there exists
minimax regret rules that satisfy a symmetry condition. Subsection
\ref{algorithm 2} specializes and leverages the insights from the general
algorithm from Section \ref{model and approximation} by incorporating
symmetry. We discuss the results from the application of the algorithm in
Subsection \ref{results} where we also consider the case where the action
space of nature is restricted by a priori information of the policymaker.
Finally in Section \ref{testing two innovations} we consider the example of
\textquotedblleft testing innovations\textquotedblright\ with more than one
alternative. The proofs of all statements are given in the Appendix. A
Supplementary Appendix contains further results on the examples in the paper.

\section{The model, its approximation, and an algorithm based on
\textquotedblleft fictitious play\textquotedblright%
\ \label{model and approximation}}

A player has to choose one of several actions $t\in\mathbb{T}:=\{1,2,...,T\}$
with randomization being allowed for. In many applications, the player is a
policymaker and $\mathbb{T}$ is the set of treatments that she considers for a population.

By $Y_{t}$ for $t\in\mathbb{T}$ we denote the random outcome if the player
chooses action $t.$ The vector of potential outcomes $(Y_{t})_{t\in\mathbb{T}%
}$ is assumed to be an element in the set $S$. In several examples in the
literature, $S=\{0,1\}^{T}$ or $S=[0,1]^{T},$ that is, the case, where the
range of outcomes is the same under every action $t\in\mathbb{T}$, and where
in the first case outcomes can be interpreted as success or failure. The
algorithm we propose applies to the case where $S=\{0,1\}^{T}.$

The setup can be interpreted as a game of the player against an antagonistic
nature that chooses a joint distribution $s\in\mathbb{S}$ for $(Y_{t}%
)_{t\in\mathbb{T}},$ a so-called \textquotedblleft state of the
world\textquotedblright, where $\mathbb{S}$ denotes the set of possible
distributions on $S.$ Oftentimes, the player is assumed to care about expected
outcome. For $t\in\mathbb{T}$ define $\mu_{t}=E_{s}Y_{t}.$

Before choosing an action, the player observes a sample $w_{N}=(t_{i}%
,y_{t_{i},i})_{i=1,...,N}$ of size $N$ with $y_{t_{i},i},$ for given $t_{i},$
denoting an independent draw from $Y_{t_{i},i}$ for $i=1,...,N$ (generated
from a certain $s\in\mathbb{S)}$. As in Stoye (2009) several \emph{sampling
designs} are possible. For example, under \emph{fixed assignment}, assume
integers $N_{t}$ for $t\in\mathbb{T}$ are given and the sample $w_{N}$
consists of $N_{t}$ independent observations under each action $t\in
\mathbb{T}$ with the sum of the $N_{t}$ equal to $N.$ Under \emph{random
assignment}, $P(t_{i}=t)=p_{t}$ for all $i=1,...,N$ and $t\in\mathbb{T},$ for
some probabilities $p_{t}$ whose sum over $t$ equals 1. Lastly, consider the
case of \emph{testing innovations}, where the policymaker knows the mean
outcome $\mu_{T}$ of treatment $T$ of a status quo treatment. She needs to
consider whether to switch from the status quo to one of $T-1$ alternative
treatments whose mean outcome is unknown. The sample $w_{N}=(t_{i},y_{t_{i}%
,i})_{i=1,...,N}$ consists of $N_{t}$ independent observations under each
action $t\in\mathbb{\{}1,...,T-1\mathbb{\}}$ with the sum of the $N_{t}$ for
$t=1,...,T-1\mathbb{\ }$equal to $N.$ Or alternatively one could generate the
sample via random assignment.

Wlog we can assume that the marginal distributions of $Y_{t}$ are independent
of each other. Then, in the case where $S=\{0,1\}^{T}$ the joint distribution
of $(Y_{t})_{t\in\mathbb{T}}$ is fully pinned down by the mean vector
$\mu=(\mu_{1},...,\mu_{T})\in\lbrack0,1]^{T}.$ Thus, from now on, in this
case, an action by nature (rather than being presented as a joint DGP for the
potential outcomes) is identified with picking a mean vector $\mu\in
\lbrack0,1]^{T}.$ The policymaker may have certain a priori knowledge that
allows her to restrict $\mu$ to a subset $M\subset\lbrack0,1]^{T}.$

We next define the notion of a \emph{treatment rule}. The task for the player
is to choose a vector $\delta(w_{N})\in\lbrack0,1]^{T}$ for every possible
sample realization $w_{N},$ whose components add up to one and represent the
probabilities with which action $t$ is chosen. Namely, which treatment the
player ultimately assigns is determined as an independent draw from the
discrete random variable $B=B(\delta(w_{N}))\in\mathbb{T}$ that equals $t$
with probability $\delta_{t}(w_{N})$ for $t\in\mathbb{T},$ where $\delta
_{t}(w_{N})$ denotes the $t$-th component of $\delta(w_{N}).$

For a given sampling design, denote by $\mathbb{D}$ the set of rules $\delta$
that the policymaker can choose from. Unless otherwise specified, $\mathbb{D}$
is taken as the set of all (measurable)\emph{ }mappings
\begin{equation}
w_{N}\mapsto\delta(w_{N})\in\Delta^{T-1}:=\{x=(x_{1},...,x_{T})^{\prime}%
\in\lbrack0,1]^{T},\text{ }x_{t}\geq0\text{ for }t=1,...,T,\text{ }%
{\textstyle\sum\nolimits_{t=1}^{T}}
x_{t}=1\} \label{set of policy rules}%
\end{equation}
with $\delta(w_{N})$ defining a probability distribution over the set
$\mathbb{T}.$ Note that we are therefore including randomized rules. In
certain applications, $\mathbb{D}$ may be restricted. For example, Ishigara
and Kitagawa (2024) use only nonrandomized linear aggregation rules to gain
tractability. In empirical applications, policy restrictions may rule out
certain elements in $\Delta^{T-1}$ from consideration.

For given $\delta\in\mathbb{D}$ and $s\in\mathbb{S}$ denote by $u(\delta,s)$
the payoff for the player. In this paper, $u(\delta,s)=E_{s}Y_{B(\delta
(w_{N}))}$ i.e. the player cares about the expected outcome.\footnote{Another
possibility is$\ u(\delta,s)=q_{s,\alpha}(Y_{B(\delta(w))}),$where
$q_{s,\alpha}(Y_{B(\delta(w))})$ denotes the $\alpha$-quantile of
$Y_{B(\delta(w))}$ when the state of the world is $s\in\mathbb{S},$ see Manski
(1988).} Given a choice $u(\delta,s)$ we use the optimality concept of
\emph{minimax regret} to choose a rule for the player. In particular, define
regret of a rule $\delta$ when the state of nature is $s$ as%
\begin{equation}
R(\delta,s)=\max_{d\in\mathbb{D}}u(d,s)-u(\delta,s). \label{regret}%
\end{equation}
The objective is to find a rule $\delta^{\ast},$ if one exists, that
satisfies
\begin{equation}
\delta^{\ast}\in\arg\min_{\delta\in\mathbb{D}}\max_{s\in\mathbb{S}}%
R(\delta,s). \label{minimax rule}%
\end{equation}
For the remainder of this paper, we assume that the player is concerned about
expected outcome. Minimax regret rules have been derived analytically only in
very few cases. We propose an algorithm to generate a treatment rule whose
maximal regret gets arbitrary close to the minimax regret value. The algorithm
is based on a result in Robinson (1951).\medskip

\textbf{Robinson's (1951) algorithm}

Robinson (1951) considers a finite two-player zero-sum game with payoff matrix
$A=(a_{ij})$ for $i=1,..,f$ and $j=1,...,g.$ When the row-player (player 1
from now on) chooses action $i$ and the column player (player 2 from now on)
chooses action $j$ then player 1 gets $a_{ij}$ and player 2 gets $-a_{ij}.$
Each player can choose a mixed strategy over her $f$ and $g$, respectively,
actions, that is, player 1 can choose a probability distribution $p_{i},$
$i=1,..,f,$ such that $p_{i}\geq0$ and $%
{\textstyle\sum\nolimits_{i=1}^{f}}
p_{i}=1$ (and $q_{j},$ $j=1,..,g,$ such that $q_{j}\geq0$ and $%
{\textstyle\sum\nolimits_{j=1}^{g}}
q_{j}=1$ for player 2$,$ respectively.) By the minimax theorem there exists
distributions such that the inequality (that holds for all distributions)
\begin{equation}
\min_{j}%
{\textstyle\sum\nolimits_{i=1}^{f}}
a_{ij}p_{i}\leq\max_{i}%
{\textstyle\sum\nolimits_{j=1}^{g}}
a_{ij}q_{j} \label{minmax inequality}%
\end{equation}
holds as an equality. Robinson (1951) provides an iterative procedure where in
the $n$-th iteration, mixed strategies $\delta^{n}$ and $\nu^{n}$ for players
1 and 2, defined by $(p_{in})_{i=1,..,f}$ and $(q_{jn})_{j=1,...,g},$
respectively, are picked such that
\begin{equation}
\lim_{n\rightarrow\infty}\min_{j}%
{\textstyle\sum\nolimits_{i=1}^{f}}
a_{ij}p_{in}=\lim_{n\rightarrow\infty}\max_{i}%
{\textstyle\sum\nolimits_{j=1}^{g}}
a_{ij}q_{jn} \label{limit is value}%
\end{equation}
and the common limit equals the value of the game. The procedure works as
follows. Start with any one of the $f$ possible actions of player 1. The mixed
strategy $\delta^{1}$ puts weight one on that action. Pick any one of the $g$
possible actions of player 2 that is a best response to $\delta^{1}$. The
distribution $\nu^{1}$ puts weight one on that action. For the iteration,
assume $\delta^{n}$ defined by $(p_{in})_{i=1,..,f}$ and $\nu^{n}$ defined by
$(q_{jn})_{j=1,...,g}$ are given. Pick any one of the $f$ possible actions of
player 1 that is a best response to $\nu^{n}$ i.e. maximizes player 1's
payoff. The mixed strategy $\delta^{n+1}$ is then defined as $\frac{n-1}%
{n}\delta^{n}+\frac{1}{n}I(\delta_{BR}^{n}),$ where $\delta_{BR}^{n}$ denotes
a best response by player 1 and $I(\cdot)$ denotes a point mass on the action
in the brackets. Pick any one of the $g$ possible actions of player 2 that is
a best response to $\delta^{n+1}$, i.e. maximizes player 2's payoff. The
distribution $\nu^{n+1}$ is defined as $\frac{n-1}{n}\nu^{n}+\frac{1}{n}%
I(\mu_{BR}^{n}),$ where $\mu_{BR}^{n}$ denotes a best response by player 2.
Theorem 1 by Robinson (1951) establishes that this algorithm guarantees that
(\ref{limit is value}) holds. Furthermore, Fudenberg and Levine (1998,
Proposition 2.2) establish that if the distributions used in each round
converge for both players, then the limits must constitute a mixed strategy
Nash equilibrium.\footnote{This procedure may not work for games with generic
payoffs or more than two players, because the empirical distribution of the
strategies may not converge. See references in Fudenberg and Levine
(1998).}\medskip

We first consider the binary case where $S=\{0,1\}^{T}$ and $\mu=(\mu
_{1},...,\mu_{T})\in M\subset\lbrack0,1]^{T}.$ The policymaker and nature play
the roles of players 1 and 2, respectively, in Robinson's (1951) setup above.
The elements $a_{ij}$ in the payoff matrix $A$ represent negative regret when
the policymaker and nature choose actions $i$ and $j,$ respectively. Player 1
aims to maximize negative regret while player 2 aims to minimize it.

In the next subsection we suggest discretizing nature's parameter space to
conform with Robinson's (1951) requirement of only finitely many actions for
both players. Given outcomes are assumed to be in $\{0,1\}$ and given there
are $T$ possible treatments on which data might be observed, there are only
finitely many different samples $w_{N}$ that can arise. Denote by $W$ the
number of different samples, $(w_{N1},...w_{NW})$ say, that can arise. Any
treatment policy $\delta$ can then by summarized as a $T\times W$ matrix (or a
$TW$ vector), where the $w$-th column contains $\delta(w_{Nw})\in\Delta^{T-1}$
for $w=1,...,W.$ If one takes as the actions for the policymaker the set of
nonrandomized treatment rules then this results in a finite set of
actions\footnote{For example, if the sample contains $N_{t}$ observations for
treatment $t$, $t\in\mathbb{T}$ then $W=2^{N}$, where $N=N_{1}+...+N_{T}.$and
there are $T^{W}$ many different nonrandomized treatment rules. That number
grows very fast in both $N$ and $T.$} and mixing these actions one obtains all
treatment rules as defined in (\ref{set of policy rules}). In some
applications the finite set of actions may contain also randomized rules. In
all applications considered below one needs to choose a finite set of actions,
denoted by $D,$ such that, when mixing over the actions in $D$ is allowed for,
then one obtains the set of all treatment rules $\mathbb{D}$ the policymaker
can choose from, that is
\begin{equation}
\mathbb{D}=\{%
{\textstyle\sum\nolimits_{i=1}^{f}}
p_{i}\delta_{i},\text{ for all }i=1,...,f,\text{ }\delta_{i}\in D,\text{
}p_{i}\geq0,\text{ and }%
{\textstyle\sum\nolimits_{i=1}^{f}}
p_{i}=1\}. \label{relation D and D}%
\end{equation}
The algorithm applies straightforwardly to any such set of action $D$ and
therefore policy restrictions can be easily implemented.

\subsection{Approximation via discretization\label{discretization}}

In this subsection we define and discuss an $\varepsilon$-discretization of
the action space of nature. Denote by $||\cdot||$ the Euclidean norm.

\textbf{Definition (}$\varepsilon$-discretization$\mathbf{)}$\textbf{.}
\emph{We call a finite subset }$M_{\varepsilon}\subset M$ \emph{an
}$\varepsilon$\emph{-discretization of the action space of nature if for any
}$\mu\in M$ \emph{we can find a }$\mu_{\varepsilon}\in M_{\varepsilon}$\emph{
such that }$\Vert\mu_{\varepsilon}-\mu\Vert<\varepsilon.$\smallskip

\textbf{Comment.} Often one can take $M_{\varepsilon}$ as $\{\mu=(\mu
_{1},...,\mu_{T})\in M;$ $\mu_{t}=\overline{n}_{t}\varepsilon$ for any
$\overline{n}_{t}\in\mathbb{N}$ for all $t=1,...,T\}$ i.e. one takes an
evenly-spaced grid with step size $\varepsilon.$ However, such a construction
will not always lead to an $\varepsilon$-discretization, for example, for
certain choices of $M$ the resulting set $M_{\varepsilon}$ could even be empty.

We now establish that as $\varepsilon\rightarrow0$ the minimax regret value of
the game in which nature is restricted to DGPs from an $\varepsilon
$-discretization $M_{\varepsilon}$ converges to the minimax regret value of
the original game. We assume that the regret function $R(\delta,\cdot)$ is
continuous in $\mu\in M$ uniformly in $\delta\in\mathbb{D},$ i.e.
$\forall\lambda>0$ there exists $\eta>0$ such that when $\Vert\mu-\mu^{\prime
}\Vert<\eta$ for $\mu,\mu^{\prime}\in M$ then for all $\delta\in\mathbb{D}$ we
have $|R(\delta,\mu)-R(\delta,\mu^{\prime})|<\lambda$. \smallskip

\begin{lemma}
\label{lem:finite approx} Consider the case $S=\{0,1\}^{T}$ and $\mu\in M$ for
$M$ compact$.$ Define%
\begin{equation}
V\equiv\inf_{\delta\in\mathbb{D}}\max_{\mu\in M}R(\delta,\mu)\text{ and }%
V_{m}\equiv\inf_{\delta\in\mathbb{D}}\max_{\mu\in M_{\varepsilon_{m}}}%
R(\delta,\mu), \label{defs value}%
\end{equation}
the values of the original and discretized games, respectively. Assume the
regret function $R(\delta,\cdot)$ is continuous in $\mu$ uniformly in
$\delta.$ Then, $\varepsilon_{m}\rightarrow0$ as $m\rightarrow\infty$ implies

$($i$)$ $V_{m}\rightarrow V$ and

$($ii$)$ $\max_{\mu\in M}R(\delta_{\varepsilon_{m}},\mu)\rightarrow V$ for any
rule $\delta_{\varepsilon_{m}}$ that satisfies $V_{m}-\max_{\mu\in
M_{\varepsilon_{m}}}R(\delta_{\varepsilon_{m}},\mu)\rightarrow0.$
\end{lemma}

\textbf{Comments.} 1. Continuity of $R(\delta,\cdot)$ is a weak assumption
that we verify in several examples below. Together with compactness of $M$ the
condition implies that $\max_{\mu\in M}R(\delta,\mu)$ exists for any
$\delta\in\mathbb{D}.$ Note that in Lemma \ref{lem:finite approx} we allow for
the possibility that a minimax regret rule may not exist. We define the value
$V$ using the infimum over all treatment rules. Often the $\inf$ in the
definition of $V$ (and $V_{m}$) can be replaced by a $\min$, e.g. if
$\max_{\mu\in M}R(\delta,\mu)$ depends continuously on $\delta$ and
$\mathbb{D}$ is compact.

2. Lemma \ref{lem:finite approx}(i) states that the minimax regret value for
the case where nature's parameter space equals $M_{\varepsilon_{m}}$ converges
to the value of the game when the parameter space equals $M$ as $\varepsilon
_{m}\rightarrow0.$ Lemma \ref{lem:finite approx}(ii) states that if one uses a
rule $\delta_{\varepsilon_{m}}$, that is approximately minimax regret in the
case where nature's parameter space is $M_{\varepsilon_{m}},$ in a situation
where nature's parameter space equals $M$ then the resulting minimax regret
value converges to the value of the game with parameter space $M$ as
$\varepsilon_{m}\rightarrow0.$ That is, $\delta_{\varepsilon_{m}}$ is
\textquotedblleft asymptotically\textquotedblright\ minimax regret in the case
where nature's parameter space equals $M$ in the sense that maximal regret for
the rule $\delta_{\varepsilon_{m}}$ converges to the minimax regret value as
one lets the \textquotedblleft stepsize\textquotedblright\ $\varepsilon_{m}$
in the discretization go to zero. We are not claiming that the rule
$\delta_{\varepsilon_{m}}$ converges to a minimax regret rule in the original
model as $\varepsilon_{m}\rightarrow0$, the convergence statement is about
maximal regret$.$ In this paper, we are proposing an algorithm that produces a
treatment rule whose maximal regret with parameter space equal to
$M_{\varepsilon_{m}}$ converges to the minimax regret value of the discretized
model (which in turn converges to the minimax regret value of the original
model as $\varepsilon_{m}\rightarrow0$.)

3. We make the discretization of nature's action space explicit in our
discussion. But note that any numerical approach to approximate minimax regret
rules via an iterative algorithm that relies on repeatedly calculating best
responses by nature through grid search also effectively uses discretization.

4. We are currently working on explicit upper bounds for $V-V_{m}$ as a
function of $\varepsilon_{m}.$ Such bounds can be worked out in our lead
examples below as expressions of Lipschitz constants of certain functions.

\subsection{Algorithm and its convergence properties\label{algorithm}}

In this subsection we describe in detail the algorithm based on Robinson
(1951) used to approximate minimax regret rules in the discretized setup, that
is, we consider the case where $S=\{0,1\}^{T}$ and $\mu=(\mu_{1},...,\mu
_{T})\in M_{\varepsilon}$ for a small $\varepsilon>0.$ The discretization
$M_{\varepsilon}$ of $M$ is kept fixed throughout the iterations. We also
allow for a possible modification suggested by Leslie and Collins (2006)
regarding the weights used in the updating of the strategies. Recall that as
specified in (\ref{relation D and D}) we denote by $D$ a finite set of
actions, such that, when mixing over the actions in $D$ is allowed for, then
one obtains the set of all treatment rules $\mathbb{D}$ the policymaker can
choose from. For example, $D$ could be chosen as the set of nonrandomized
treatment rules which is finite and satisfies (\ref{relation D and D}).

The algorithm provides a sequence of treatment rules and strategies by nature.
For $n\geq1$ we denote by $\delta^{n}$ the treatment rule in the $n$-th
iteration, by $\nu^{n}$ the empirical distribution of strategies for nature up
to time $n,$ and by $R^{n}$ a lower bound on the minimax regret value obtained
in the $n$-th iteration. To apply the algorithm, pick a small threshold
$\xi>0$ and a sequence of updating weights $\alpha_{n}$ for $n=1,2,...$ that
will be specified further below$.$

\begin{algorithm}
\label{general algorithm}

\textbf{Initialization: }

i$)$ Define $\delta^{1}$ as one of the actions in $D$.

ii$)$ Initialize $\nu^{0}=0$ and $R^{0}=0.$

\textbf{Iteration. For }$n=1,2,3...$\textbf{ DO:}

i$)$ Find a best response $\mu_{BR}^{n}\in M_{\varepsilon}$ by nature to
$\delta^{n}$ that is, a mean vector that maximizes regret given $\delta^{n}:$
\begin{equation}
R(\delta^{n},\mu_{BR}^{n})=\max_{\mu\in M_{\varepsilon}}R(\delta^{n},\mu).
\label{best response for nature}%
\end{equation}

ii$)$ If
\begin{equation}
R(\delta^{n},\mu_{BR}^{n})-R^{n-1}<\xi\label{STOPPING RULE}%
\end{equation}
then \textbf{break. }Use the rule $\delta^{n}.$

iii$)$ Update nature's mixed strategy to be a weighted average of strategies
used up to this point, i.e.
\begin{equation}
\nu^{n}=(1-\alpha_{n})\nu^{n-1}+\alpha_{n}I(\mu_{BR}^{n}),
\label{UPDATE NATURE}%
\end{equation}
where $I(x)$ denotes a point mass of size 1 at the point $x\in\lbrack0,1]^{T}$.

iv$)$ Compute a best response $\delta_{BR}^{n}$ in $D$ by the policymaker to
$\nu^{n}$, that is,
\begin{equation}
R^{n}:=R(\delta_{BR}^{n},\nu^{n})=\min_{\delta\in D}R(\delta,\nu^{n}).
\label{UPDATE LOWER VALUE}%
\end{equation}
v$)$ Update the treatment rule by%
\begin{equation}
\delta^{n+1}=(1-\alpha_{n+1})\delta^{n}+\alpha_{n+1}\delta_{BR}^{n}.
\label{UPDATE PLAYER}%
\end{equation}

\end{algorithm}

\textbf{Comments. }1. Regarding the iteration step iv) note that for a best
response $\delta_{BR}^{n}$ against $\nu^{n}$ it is necessary and sufficient
that whenever
\begin{equation}
E(Y_{t}|w_{N})>E(Y_{t^{\prime}}|w_{N})\text{ for }t,t^{\prime}\in\mathbb{T}
\label{condition 0}%
\end{equation}
holds, $\delta_{BR}^{n}$ gives zero weight to the \textquotedblleft
dominated\textquotedblright\ treatment $t^{\prime}.$\footnote{One can always
find a nonrandomized best response $\delta_{BR}^{n}.$ Unless of course one
imposes certain restrictions on the class of treatment rules that impose
randomization.} Importantly, an application of \textbf{Bayes' rule} then
allows for a straightforward solution to the minimization problem in iv) that
provides a very significant computational gain relative to other methods, such
as grid search, to solve that problem. Namely, assume nature picks the mixed
strategy $\nu^{n}$ that puts nonzero weights $p_{m}$ on exactly the mean
vectors $\overline{\mu}_{m}=(\mu_{1m},...,\mu_{Tm})\in M_{\varepsilon}$ for
$m=1,...,\overline{m}.$ Then,%
\begin{align}
&  E(Y_{t}|w_{N})\nonumber\\
&  \overset{}{=}P(Y_{t}\overset{}{=}1|w_{N})\nonumber\\
&  \overset{}{=}%
{\textstyle\sum\nolimits_{m=1}^{\overline{m}}}
P(Y_{t}\overset{}{=}1\text{ \& nature picks }\overline{\mu}_{m}|w_{N}%
)\nonumber\\
&  \overset{}{=}%
{\textstyle\sum\nolimits_{m=1}^{\overline{m}}}
P(Y_{t}\overset{}{=}1|\text{nature picks }\overline{\mu}_{m}\text{ \& }%
w_{N})P(\text{nature picks }\overline{\mu}_{m}|w_{N})\nonumber\\
&  \overset{}{=}%
{\textstyle\sum\nolimits_{m=1}^{\overline{m}}}
P(Y_{t}\overset{}{=}1|\text{nature picks }\overline{\mu}_{m})P(w_{N}%
|\text{nature picks }\overline{\mu}_{m})p_{m}/P(w_{N})\nonumber\\
&  \overset{}{=}%
{\textstyle\sum\nolimits_{m=1}^{\overline{m}}}
\mu_{tm}P(w_{N}|\text{nature picks }\overline{\mu}_{m})p_{m}/P(w_{N}),
\label{derivation for bayes}%
\end{align}
and therefore (\ref{condition 0}) holds iff
\begin{equation}%
{\textstyle\sum\nolimits_{m=1}^{\overline{m}}}
\mu_{tm}P(w_{N}|\text{nature picks }\overline{\mu}_{m})p_{m}>%
{\textstyle\sum\nolimits_{m=1}^{\overline{m}}}
\mu_{t^{\prime}m}P(w_{N}|\text{nature picks }\overline{\mu}_{m})p_{m}.
\label{condition}%
\end{equation}
For a given sample $w_{N}$ and choices of $p_{m}$ and $\overline{\mu}_{m}$ for
$m=1,...,\overline{m}$ the expressions in (\ref{condition}) can be easily
calculated (where $P(w_{N}|$nature picks $\overline{\mu}_{m})$ obviously
depends on the sampling design that is being employed).

2. Choosing the weights $\alpha_{n}=1/n$ corresponds to the algorithm proposed
in Robinson (1951). In that case, every iteration involves optimizing against
the empirical distribution function of strategies chosen up to this point.
Other choices of $\alpha_{n}$ satisfying $\alpha_{n}>0,$ $\alpha
_{n}\rightarrow0$, and $\sum_{n=1}^{\infty}\alpha_{n}=\infty$ have been
suggested in Leslie and Collins (2006), e.g. $\alpha_{n}=(C+n)^{-\eta}$ for
some constants $\eta\in(0,1]$ and $C>0$ (which nests $\alpha_{n}=1/n$ when
$C=0$ and $\eta=1)$ or $\alpha_{n}=1/\log(C+n).$ Based on our experience in
several examples, we find that $\alpha_{n}=(C+n)^{-\eta}$ with $\eta=.7$ and
$C=5$ works quite well.

3. In the applications of the algorithm considered below regret goes to zero
as the sample size $N$ goes to infinity. Therefore, a \textquotedblleft
small\textquotedblright\ threshold $\xi$ in one setting may not be small in
another setting. It therefore makes sense to modify the break command in
(\ref{STOPPING RULE}) by also making sure that the approximation error
relative to the actual regret is small, that is to check whether
$(R(\delta^{n},\mu_{BR}^{n})-R^{n-1})/R^{n-1}<\xi.$

4. In practice, after the algorithm terminates, one does not necessarily use
the treatment rule obtained in the last iteration. One tracks the maximum
regret level for the treatment rules generated during the entire iteration
process, and chooses the treatment rule that has the minimum level of maximum
regret among those rules. This is because there is no guarantee that maximal
regret will decrease in each iteration, so it is possible that the rule
generated in the last iteration is not the best one amongst the rules
considered during all iterations.

5. \textbf{Bounds:} Note that $R^{n}=R(\delta_{BR}^{n},\nu^{n})$ denotes the
minimal regret against a particular mixed strategy by nature and therefore is
nonbigger than the minimax regret value. Analogously, $R(\delta^{n},\mu
_{BR}^{n})$ denotes maximal regret against a particular mixed strategy by the
policymaker and therefore is nonsmaller than the minimax regret value. If
$R^{n}=R(\delta_{BR}^{n},\nu^{n})$ and $R(\delta^{n},\mu_{BR}^{n})$ are close
to each other, they are both necessarily close to the minimax regret value of
the discretized model. In any iteration of the algorithm $R(\delta^{n}%
,\mu_{BR}^{n})-R^{n}$ provides an upper bound for how far away the maximal
regret of the currently considered treatment rule is from the minimax regret
value. \medskip

If one lets $n$ go to infinity (and never uses the break command in step ii)
of the iteration) then by Robinson's (1951, Theorem 1) in the case of
$\alpha_{n}=1/n$ and by Leslie and Collins (2006, Corollary 5) for the other
cases in the proposition below, $R^{n}$ converges to the minimax regret value
of the discretized game. We formulate this important result as a proposition.

\begin{proposition}
\label{proof of convergence}Assume the sequence $\alpha_{n}$ satisfies
$\alpha_{n}>0,$ $\alpha_{n}\rightarrow0$, and $\sum_{n=1}^{\infty}\alpha
_{n}=\infty.$ Then the maximal regret from the sequence of treatment rules
that is generated from Algorithm \ref{general algorithm}, $\max_{\mu\in
M_{\varepsilon}}R(\delta^{n},\mu),$ converges as $n\rightarrow\infty$ to the
minimax regret value of the $\varepsilon$-discretized model.
\end{proposition}

\textbf{Comment.} 1. Leslie and Collins (2006) provide some evidence that for
the more general choices of weights $\alpha_{n}$ one might be able to speed up
the convergence rate relative to the choice $\alpha_{n}=1/n.$ An important
topic for future research is to theoretically work out the convergence speed
as a function of the chosen weights.

\subsection{\textbf{Coarsening\label{coarsening subsection}}}

Assume one has found a minimax regret rule $\delta$ (or approximate minimax
regret rule $\delta_{\varepsilon})$ in the setup with $S=\{0,1\}^{T}$ and mean
vector $\mu\in M\subset\lbrack0,1]^{T}$ for a certain sampling design, such as
fixed assignment with $N_{t}$ observations for $t\in\mathbb{T}$ or random
assignment with $P(t_{i}=t)=p_{t}$ for all $i=1,...,N$ and $t\in\mathbb{T}.$
We show that then the so-called coarsening approach can be used to generate a
minimax regret rule (or approximate minimax regret rule) for the case
$S=[0,1]^{T}$ and the same restriction on the mean vector. The
\textquotedblleft coarsening trick\textquotedblright\ proceeds as follows:

Coarsen the outcomes $y_{t_{1},1},...,y_{t_{N},N}$ in the sample $w_{N}%
=(t_{i},y_{t_{i},i})_{i=1,...,N}$ by replacing each $y_{t_{i},i}$ by an
independent draw $\widetilde{y}_{t_{i},i}$ from $\widetilde{Y}_{t_{i},i}%
\in\{0,1\}$ from a Bernoulli random variable with success probability
$y_{t_{i},i}$ for $i=1,...,N.$ Denote by $\widetilde{w}_{N}=(t_{i}%
,\widetilde{y}_{t_{i},i})_{i=1,...,N\text{ }}$ the coarsened sample. Define a
new treatment rule $\delta^{C}$ by setting $\delta^{C}(w_{N})=\delta
(\widetilde{w}_{N})$ (or by $\delta_{\varepsilon}^{C}(w_{N})=\delta
_{\varepsilon}(\widetilde{w}_{N})).$

Part i) of the following proposition states that then the treatment rule
$\delta^{C}$ is minimax regret with $S=[0,1]^{T}$ with mean restriction $M.$
The coarsening approach has been introduced when $T=2$ and $M=[0,1]^{T}$, see,
e.g. Cucconi (1968), Gupta and Hande (1992), Schlag (2003, 2006), and Stoye
(2009), and applied with arbitrary $T$ in Chen and Guggenberger (2024) but it
can also be employed here with mean restriction $M$. Part ii) of the
proposition shows that coarsening can also be applied to approximate minimax
regret rules in the case $S=\{0,1\}^{T}$ and given $M$ to obtain approximate
minimax regret rules in the case $S=[0,1]^{T}$ with same mean restriction $M.$

\begin{proposition}
\label{coarsening}Assume the sample is generated by fixed, random assignment
or testing innovations.

i$)$ Assume $\delta$ is a minimax regret rule when $S=\{0,1\}^{T}$ and $\mu\in
M\subset\lbrack0,1]^{T}.$ Then the treatment rule $\delta^{C}$ defined by
$\delta^{C}(w_{N})=\delta(\widetilde{w}_{N})$ is minimax regret with
$S=[0,1]^{T}$ and $\mu\in M\subset\lbrack0,1]^{T}.$

ii$)$ Assume for some $\varepsilon>0$, $\delta_{\varepsilon}$ is an
$\varepsilon$-approximate minimax regret rule when $S=\{0,1\}^{T}$ and $\mu\in
M\subset\lbrack0,1]^{T},$ meaning, $\min_{\delta\in\mathbb{D}}\max_{\mu\in
M}R(\delta,\mu)+\varepsilon\geq\max_{\mu\in M}R(\delta_{\varepsilon},\mu).$
Then the treatment rule $\delta_{\varepsilon}^{C}$ defined by $\delta
_{\varepsilon}^{C}(w_{N})=\delta_{\varepsilon}(\widetilde{w}_{N})$ is an
$\varepsilon$-approximate minimax regret rule when $S=[0,1]^{T}$ and $\mu\in
M\subset\lbrack0,1]^{T},$ meaning, $\min_{\delta\in\mathbb{D}}\max
_{s\in\mathbb{S}\text{ s.t. }\mu\in M}R(\delta,s)+\varepsilon\geq\max
_{s\in\mathbb{S}\text{ s.t. }\mu\in M}R(\delta_{\varepsilon}^{C}%
,s).\footnote{Recall that by $\mathbb{S}$ denotes the set of possible
distributions for $(Y_{1},...Y_{T})$ on $S.$ Recall also that wlog we can
assume that the marginals $Y_{t}$ for $t\in\mathbb{T}$ are independent.}$
\end{proposition}

\textbf{Comments. }1. The proof of Proposition \ref{coarsening} i) adapts the
proofs in Schlag (2006) and Stoye (2009) from the case with equal sample
sizes, $T=2$ and $M=[0,1]^{2}$ to the case considered here with arbitrary $T$
and $M$ and the $N_{t}$ not necessarily all equal. The proposition could
likely be generalized to alternative sampling designs.

2. Proposition \ref{coarsening} ii) is an important result. It establishes
that approximate minimax regret rules can be obtained for complex problems
through a combination of a numerical approach applied in a simpler setup and
the coarsening approach. In particular, it applies to the setup of the example
considered in Section \ref{lead example}.

\section{Treatment assignment with unbalanced samples\label{lead example}}

In this section we look at an important special case of the general setup in
the previous section, namely the case of a policymaker who needs to choose
between two treatments after being informed by a sample consisting of $N_{t}$
independent observations from each treatment $t=1,2.$ That is, here, $T=2$ and
$\mathbb{T}=\{1,2\}.$ We assume that the policymaker's objective is expected
outcome, $u(\delta,s)=E_{s}Y_{B(\delta(w_{N}))}$ and that realizations are
either successes or failures, that is $S=\{0,1\}^{2}.$ We can then apply the
coarsening approach to deal with the general case $S=[0,1]^{2}.$ Given that
observations are independent wlog we can assume that under $s\in\mathbb{S}$
the marginals $(Y_{1},Y_{2})$ are independent and that therefore the joint
distribution is fully pinned down by the mean vector $(\mu_{1},\mu_{2})\in
M=[0,1]^{2}.$ We first consider the case here where the policymaker is
completely agnostic and does not have any a priori information about the mean
vector. Then, in Section \ref{restrict} we apply the algorithm to find
approximate minimax regret rules when $M$ is a strict subset of $[0,1]^{2}.$
From now on, we identify any $s\in\mathbb{S}$ with its resulting mean vector.

An analytical formula for minimax regret rules is known \emph{only} in the
case where the $N_{t}$ are equal, that is, in the case of \textquotedblleft
matched treatment assignment\textquotedblright. The following statement holds
for any integer $T\geq2.$

\begin{proposition}
\label{Analogue to Proposition 1 in Stoye 2009}Take $\mathbb{S}$ as the set of
all distributions on $\{0,1\}^{T}$. Denote by $n_{t}$ the number of successes
for treatment $t$ in the sample. Define
\begin{equation}
W=\arg\max_{t\in\mathbb{T}}n_{t} \label{n_m}%
\end{equation}
the set of all those treatments for which most successes are observed in the
sample. Define
\begin{equation}
\delta_{t}(w_{N})=1/|W|\text{ if }t\in W\text{ and }\delta_{t}(w_{N})=0\text{
otherwise,} \label{minimax Bernoulli case}%
\end{equation}
where for a set $W$ we denote by $|W|$ the number of elements in $W$ and
$\delta_{t}$ for $t\in\mathbb{T}$ denotes the $t$-th component of the
treatment rule $\delta.$ If the $N_{t}$ are all equal then $\delta
\in\mathbb{D}$ is a minimax regret treatment rule.
\end{proposition}

\textbf{Comments.} 1. Proposition
\ref{Analogue to Proposition 1 in Stoye 2009} was proven by Schlag (2006) and
Stoye (2009) for the case $T=2$ and for arbitrary finite $T$ by Chen and
Guggenberger (2024). The proof uses the well-known Nash equilibrium approach,
see e.g. Berger (1985), that is based on the insight that the action of the
policymaker in any Nash equilibrium in the zero sum game between the
policymaker and nature (with nature's payoff being regret and the
policymaker's payoff being negative regret) is automatically a minimax regret
rule. See Chen and Guggenberger (2024) for a reproduction of the simple proof
of that result. Note that in the zero sum game nature is allowed to randomize
over its actions where an action by nature is simply a vector of means
$(\mu_{1},\mu_{2})\in\lbrack0,1]^{2}.$ From now on we denote by $\Delta
\mathbb{S}$ the space of randomized rules for nature.\footnote{Note though
that being part of a Nash equilibrium is a sufficient but not a necessary
condition for a minimax rule. See Guggenberger, Mehta, and Pavlov (2024) for
an example of a minimax rule that is not part of a Nash equilibrium.}%
$\medskip$

When the $N_{t}$ are not all the same then no analytical results are known
about minimax regret rules - even in the case $T=2$. We make a tiny bit of
progress in that regard and next provide an analytical formula for minimax
regret rules in the special case where $\min_{t=1,2}\{N_{t}\}=0.$ Wlog in the
following proposition assume $N_{1}=0$ and $N_{2}>0$.

\begin{proposition}
\label{analytic result}In the case where $N_{1}=0$ and $N_{2}>0$ the treatment
rule $\delta$ defined by $\delta_{2}(0,n_{2})=n_{2}/N_{2}$ is minimax regret.
\end{proposition}

\textbf{Comments. }1. The proof establishes that a least favorable
distribution by nature is given by randomizing uniformly between the two mean
vectors $(0,1/2)$ and $(1,1/2).$

2. Note that for any $n_{2}$ and $n_{2}^{\prime}$ in $\{0,1...,N_{2}\}$ such
that $n_{2}+n_{2}^{\prime}=N_{2}$, we have $\delta_{2}(0,n_{2})+\delta
_{2}(0,n_{2}^{\prime})=(n_{2}+n_{2}^{\prime})/N_{2}=1.$ Thus, the treatment
rule satisfies the type of symmetry condition that we formally introduce below
in (\ref{symmetry def}).

3. We put substantial effort in deriving analytical derivations for minimax
regret rules in more general cases but were not successful. Therefore, in what
follows below, we will instead approximate minimax regret rules by a version
of the algorithm from Section \ref{model and approximation} based on
fictitious play.$\medskip$

In the next subsection, we show that minimax regret rules can be found in a
class of treatment rules restricted by a symmetry condition. We then provide
an algorithm that leverages the insights of fictitious play with the
restrictions imposed by symmetry, and finally report results from the
algorithm for several choices of $(N_{1},N_{2}).$

From now on, we represent a sample $w_{N}$ by the number of successes $n_{t}$
for each treatment, that is, (with some abuse of notation) we write
$w_{N}=(n_{1},n_{2}).$

\subsection{Symmetry restriction\label{symmetry}}

In the context of treatment assignment with unbalanced samples when $T=2$ and
$S=\{0,1\}^{2}$ we make the following definition.

\textbf{Definition (symmetric treatment rule).} \emph{We call a treatment rule
}$\delta$\emph{ symmetric when}
\begin{equation}
\delta_{2}(w_{N})+\delta_{2}(w_{N}^{\prime})=1 \label{symmetry def}%
\end{equation}
\emph{whenever }$w_{N}=(n_{1},n_{2})$\emph{ and }$w_{N}^{\prime}%
=(n_{1}^{\prime},n_{2}^{\prime})$\emph{ satisfy}%
\begin{equation}
n_{t}+n_{t}^{\prime}=N_{t}\text{ for }t=1,2. \label{summability}%
\end{equation}

\textbf{Comments}. 1. Note that when $\delta$ is symmetric then trivially also
$\delta_{1}(w_{N})+\delta_{1}(w_{N}^{\prime})=1$ for all $w_{N}=(n_{1},n_{2})$
and $w_{N}^{\prime}=(n_{1}^{\prime},n_{2}^{\prime})$ that satisfy
(\ref{summability}). Further note, that when both $N_{1}$ and $N_{2}$ are
even, then necessarily $\delta_{2}(w_{N})=1/2$ for $w_{N}=(N_{1}/2,N_{2}/2).$
Therefore, symmetric rules sometimes need to randomize.

2. Symmetry conditions are used in the analytical derivations of minimax
regret rules in Stoye (2009, Proposition 1) and Chen and Guggenberger (2024,
Proposition 1) but the symmetry condition defined here in the particular
context of unbalanced samples appears to be new.

3. The main purpose of the symmetry condition employed here is to gain
computational improvements. Note that if $\delta$ is symmetric then knowing
$\delta_{2}(w_{N})$ automatically fixes the value for $\delta_{2}%
(w_{N}^{\prime}).$ E.g. if $N_{1}=1$ and $N_{2}=5$ then knowing the value of
$\delta_{2}(0,n_{2})$ for $n_{2}=0,1,...,5$ pins down the value of $\delta
_{2}(1,5-n_{2}).$ Therefore rather than searching over all $2\times6=12$
values of $\delta_{2}(n_{1},n_{2})$ for $n_{1}=0,1$ and $n_{2}=0,1,...,5$ one
only needs to consider values for the 6 values of $\delta_{2}(0,n_{2}).$ If
for instance one were to employ an algorithm whose computation time increases
exponentially with $(n_{1}+1)(n_{2}+1)$ then searching in the class of
symmetric rules instead would reduce the computation time by a factor
$\exp((n_{1}+1)(n_{2}+1))/\exp(.5(n_{1}+1)(n_{2}+1)).\medskip$

We next show why symmetry is an important condition. The next proposition
establishes that one can find an action pair $(\delta,s)$ with $\delta
\in\mathbb{D}$ and $s\in\Delta\mathbb{S}$ for the policymaker and nature,
respectively, that constitutes a Nash equilibrium and where $\delta$ is symmetric.

\begin{proposition}
\label{symmetry statement}$($i$)$ Suppose nature plays a mixed strategy that
picks the mean vectors $\overline{\mu}_{m}=(\mu_{1m},\mu_{2m})$ with
probability $p_{m}>0$ for $m=1,...,2\overline{m}$ for some $\overline{m}%
\geq1,$ where $p_{1}+...+p_{2\overline{m}}=1$ and for all $m=1,...,\overline
{m}$ we have
\begin{equation}
(\mu_{1m},\mu_{2m})=(1-\mu_{1m+\overline{m}},1-\mu_{2m+\overline{m}})\text{
and }p_{m}=p_{m+\overline{m}}. \label{cond sym}%
\end{equation}
Then if there exists a treatment rule $\delta$ that is a best response to that
strategy $($defined by $\overline{\mu}_{m}$ and $p_{m}$ for
$m=1,...,2\overline{m})$ then it can be chosen to satisfy $($%
\ref{symmetry def}$)$.

$($ii$)$ Assume the policymaker chooses a symmetric treatment rule $\delta$
and $(\mu_{1},\mu_{2})$ is a best response by nature. Then also $(1-\mu
_{1},1-\mu_{2})$ is a best response.
\end{proposition}

\textbf{Comments. }1. In the setup of Proposition \ref{symmetry statement}(i)
nature randomizes over mean vectors that come in pairs. For every mean vector
$(\mu_{1m},\mu_{2m})$ that nature picks with positive probability the mean
vector $(1-\mu_{1m},1-\mu_{2m})$ is also picked with the same probability.
Together with Proposition \ref{symmetry statement}(ii) it follows that a Nash
equilibrium of the unrestricted game exists in which the policymaker plays a
symmetric treatment rule. The proof of Proposition \ref{symmetry statement}%
(ii) establishes that if $\delta$ is symmetric then $R(\delta,(1-\mu_{1}%
,1-\mu_{2}))=R(\delta,(\mu_{1},\mu_{2})).$

2. Proposition \ref{symmetry statement} implies that if there exists a minimax
regret rule that is part of a Nash equilibrium then
\begin{equation}
\inf_{\delta\in\mathbb{D}}\max_{\mu\in\lbrack0,1]^{2}}R(\delta,\mu
)=\inf_{\substack{\delta\in\mathbb{D}\\\delta\text{ is symmetric}}%
}\max_{\substack{s\in\Delta\mathbb{S}\\s\text{ satisfies (\ref{cond sym})}%
}}R(\delta,\mu). \label{alternative minimax characterization}%
\end{equation}
A mixed strategy $s\in\Delta\mathbb{S}$ by nature is a random vector with
outcomes in $[0,1]^{2}$ that specifies which mean vector nature chooses. The
condition (\ref{cond sym}) restricts this random variable to only pick a
certain number of pairs $(\mu_{1},\mu_{2})$ and $(1-\mu_{1},1-\mu_{2})$ whose
two elements have to be chosen with equal probabilities. In the next
subsection we are applying Algorithm \ref{general algorithm} to the
formulation of the problem on the right side of the equality in
(\ref{alternative minimax characterization}) using a discretized version of
nature's parameter space.

3. The symmetry condition is relevant in the case where $M$ satisfies the
following restriction. If
\begin{equation}
(\mu_{1},\mu_{2})\in M\text{ then }(1-\mu_{1},1-\mu_{2})\in M.
\label{sym in M}%
\end{equation}
That restriction is satisfied in the lead example where $M=[0,1]^{2}.$

4. An important question for current research concerns the development of
techniques that allow one to systematically determine the set of all symmetry
conditions in this and any other application. We came across the above
symmetry condition after we applied the algorithm for small sample sizes and
noted a pattern in the obtained approximation of minimax regret rules. We
don't know whether there are other symmetry conditions that we have not yet discovered.

\subsection{Algorithm\label{algorithm 2}}

To justify the discretization of nature's action space below, we first
establish that in the example considered here, the assumption of Lemma
\ref{lem:finite approx} about the regret function $R(\delta,\cdot)$ being
continuous in $\mu\in M$ uniformly in $\delta\in\mathbb{D}$ is
satisfied.\smallskip

\begin{lemma}
\label{lem:continuity in lead example}$\forall\lambda>0$ there exists $\eta>0$
such that when $\Vert\mu-\mu^{\prime}\Vert<\eta$ for $\mu,\mu^{\prime}\in M$
then for all $\delta\in\mathbb{D}$ we have $|R(\delta,\mu)-R(\delta
,\mu^{\prime})|<\lambda$.
\end{lemma}

For any $\varepsilon=p^{-1}$ for $p\in\mathbb{N}$ we will use the set
\begin{equation}
M_{\varepsilon}=\{(\mu_{1},\mu_{2});\mu_{j}=p_{j}/p\text{ for some }p_{j}%
\in\{0,1,...,p\}\text{ for }j=1,2\} \label{eps discretization}%
\end{equation}
as the $\varepsilon$-discretization of the set $(\mu_{1},\mu_{2})\in
M=[0,1]^{2}.$ That guarantees that if $(\mu_{1},\mu_{2})\in M_{\varepsilon}$
then also $(1-\mu_{1},1-\mu_{2})\in M_{\varepsilon}.$

Next, we apply the general Algorithm \ref{general algorithm} from Subsection
\ref{algorithm} to the problem considered here, assuming that (\ref{sym in M})
holds. Using the insights from Proposition \ref{symmetry statement} as
expressed in (\ref{alternative minimax characterization}) we exploit the
particular symmetrical structure of the current application to simplify the
complexity of the problem. Namely, we take $D$ as the finite set of symmetric
rules with outcomes in $\{0,1\}$ except for samples $w_{N}=(n_{1},n_{2})$ for
which $n_{1}=N_{1}/2$ and $n_{2}=N_{2}/2$ (which can only happen when both
sample sizes are even). For nature, the finite set of actions, $M_{\varepsilon
}^{Na}$ say, consists of those mixed strategies that uniformly randomize
between a pair of $(\mu_{1},\mu_{2})$ and $(1-\mu_{1},1-\mu_{2})$ with
$(\mu_{1},\mu_{2})\in M_{\varepsilon}$. The algorithm then follows the exact
same steps as in the Algorithm \ref{general algorithm} but with the particular
choices of $D$ and finite set of actions for nature as just described. For
clarity, we write down the entire algorithm again.

\begin{algorithm}
\label{specialized algorithm}

\textbf{Initialization: }

i$)$ For $\delta^{1}$ use the following \textquotedblleft empirical success
rule,\textquotedblright\ $\delta_{2}^{1}(n_{1},n_{2})=1$ if $n_{1}/N_{1}%
<n_{2}/N_{2},$ $\delta_{2}^{1}(n_{1},n_{2})=0$ if $n_{1}/N_{1}>n_{2}/N_{2},$
$\delta_{2}^{1}(n_{1},n_{2})=1$ if $n_{1}/N_{1}=n_{2}/N_{2}>1/2$, $\delta
_{2}^{1}(n_{1},n_{2})=0$ if $n_{1}/N_{1}=n_{2}/N_{2}<1/2,$ and $\delta_{2}%
^{1}(n_{1},n_{2})=1/2$ otherwise $($that is, when $n_{1}/N_{1}=n_{2}%
/N_{2}=1/2).$

ii$)$ Initialize $\nu^{0}=0$ and $R^{0}=0.$

\textbf{Iteration: For }$n=1,2,3...$ \textbf{DO}:\textbf{ }

i$)$ Find a best response $\mu_{BR}^{n}\in M_{\varepsilon}^{Na}$ to
$\delta^{n}$ for nature, that is, a mixed strategy with equal weights for
$\mu^{n}$ and $(1,1)-\mu^{n}$ for a $\mu^{n}\in M_{\varepsilon}.$ Let
\begin{equation}
R(\delta^{n},\mu_{BR}^{n})=\max_{\mu\in M_{\varepsilon}^{Na}}R(\delta^{n}%
,\mu). \label{regret updated}%
\end{equation}

ii$)$ If
\begin{equation}
R(\delta^{n},\mu_{BR}^{n})-R^{n-1}<\xi
\end{equation}
then \textbf{break. }Use the rule $\delta^{n}.$

iii$)$ Update nature's mixed strategy to be the following weighted average
\begin{equation}
\nu^{n}=(1-\alpha_{n})\nu^{n-1}+(\alpha_{n}/2)(I(\mu^{n})+I((1,1)-\mu^{n})),
\label{update nature example}%
\end{equation}
where $I(x)$ denotes a point mass of size 1 at the point $x\in\lbrack0,1]^{2}$.

iv$)$ Compute a best response $\delta_{BR}^{n}\in D$ by the policymaker to
$\nu^{n}$, that is,
\begin{equation}
R^{n}:=R(\delta_{BR}^{n},\nu^{n})=\min_{_{\delta\in D}}R(\delta,\nu^{n}).
\end{equation}
v$)$ Update the treatment rule by%
\begin{equation}
\delta^{n+1}=(1-\alpha_{n+1})\delta^{n}+\alpha_{n+1}\delta_{BR}^{n}.
\end{equation}

\end{algorithm}

\textbf{Comment.} 1. To implement the algorithm, one has to chose
$p=1/\varepsilon\in\mathbb{N}.$ That choice, obviously, is a trade-off between
computational effort and closeness of the minimax regret value in the original
model and the $\varepsilon$-discretized one. By increasing $p$ nature becomes
\textquotedblleft more powerful\textquotedblright\ and thus the minimax regret
value in the $\varepsilon$-discretized model increases in $p$ and (as shown in
Lemma \ref{lem:finite approx}) converges to the minimax regret value of the
original model as $p\rightarrow\infty.$ Each iteration step i) is solved by a
grid search over all $(p+1)^{2}$ possible choices of mean vectors.

2. Note that $\delta^{1}$ satisfies the symmetry condition. It is randomized
when both $N_{1}$ and $N_{2}$ are even for samples $w_{N}$ such that
$n_{1}/N_{1}=n_{2}/N_{2}=1/2.$ Alternatively, one could initiate the algorithm
with any other symmetric treatment rule. We suggest picking the empirical
success rule because it is a good starting point as it is known that its
maximal regret converges to zero at the optimal rate, see Kitagawa and Tetenov
(2018). Note that$\ \nu^{n}$ satisfies the condition given in the Proposition
\ref{symmetry statement}(i).

3. The best response of the policymaker is calculated via Bayes rule as in
(\ref{condition}) and uses the formula
\begin{equation}
P(w_{N}|\text{nature picks }\overline{\mu}_{m})=%
{\textstyle\prod\nolimits_{t=1}^{2}}
\binom{N_{t}}{n_{t}}\mu_{tm}^{n_{t}}(1-\mu_{tm})^{N_{t}-n_{t}}
\label{bayes in example}%
\end{equation}
when $w_{N}=(n_{1},n_{2}).$

4. Given this is a special case of the general setup, Proposition
\ref{proof of convergence} about convergence of the maximal regret of the
sequence of treatment rules to the minimax regret value of the $\varepsilon
$-discretized model continues to hold.

5. When potential outcomes are elements of $[0,1]$ rather than $\{0,1\}$ one
applies the coarsening approach to the obtained treatment rule from Algorithm
\ref{specialized algorithm}.

6. If there are discrete covariates $X$ in the model then an important result
in Stoye (2009) states that one obtains a minimax regret rule by simply
solving the conditional-on-$X$ problems.

\subsection{Results\label{results}}

In this subsection, we investigate the performance of Algorithm
\ref{specialized algorithm} for the example of unbalanced samples with $T=2.$

We first consider the case of \textbf{balanced samples} where $N_{1}=N_{2}.$
The reason is that in that case Proposition
\ref{Analogue to Proposition 1 in Stoye 2009} provides the minimax regret rule
in analytical form and Stoye (2009, Corollary 1) provides a formula to
calculate the minimax regret value. The latter allows us to directly assess
the performance of the algorithm. We also use the case of balanced samples to
experiment with different weighting schemes $\alpha_{n}$ and initializations
of the treatment rule.

Specifically, we consider $N_{1}=N_{2}\in\{5,10,20,40,60,80,100,200\}$. In the
discretization of the parameter space for nature in (\ref{eps discretization})
we use $p=1000$ throughout.\footnote{We also do some robustness checks by
comparing the results to the case where we only take $p=500$ and the results
vary only very slightly$.$ In Stoye (2009, Corollary 1) we use stepsize
.000001 in a gridsearch over different values of $a$ (that appears in Stoye
(2009, Corollary 1) to evaluate the formula for minimax regret given there.}

First we experiment with different choices $\eta$ and $C$ in the weighting
scheme $\alpha_{n}=(C+n)^{-\eta}$ discussed in Leslie and Collins (2006) when
$N_{1}=N_{2}=40$. Specifically, we consider a grid of values for $\eta
\in\lbrack.1,1]$ and for $C\in\lbrack0,10].$ Note that the particular choice
$C=0$ and $\eta=1$ leads to the weights proposed in Robinson (1951)$.$%
\footnote{Because $\delta^{n+1}=(1-\alpha_{n+1})\delta^{n}+\alpha_{n+1}%
\delta_{BR}^{n}.$one obtains $\delta^{1}=\delta_{BR}^{1}$ with $\alpha
_{n}=1/n$ and the impact of the initialization is only transferred through its
impact on $\mu_{BR}^{n}.$} This experiment suggests that $\eta\in
\lbrack.5,.7]$ and $C\in\lbrack5,10]$ work \textquotedblleft
best.\textquotedblright\ We will only report results for $\alpha_{n}=1/n$ and
$\alpha_{n}=(5+n)^{-.7}$ below. Those results will document the importance
that the weighting scheme has on the convergence speed of maximal regret to
the minimax regret value.

We also experiment with two different initializations $\delta^{1}$ of the
algorithm, namely, we initialize with the particular empirical success (ES)
rule suggested in Algorithm \ref{specialized algorithm}. However, given that
this initialization is very close to the actual minimax regret rule in the
case of balanced samples, we also consider one that is very different from the
actual minimax regret rule to challenge the algorithm. Namely we take the
symmetric $\delta^{1}$ defined by $\delta_{2}^{1}(n_{1},n_{2})=1$ if
$n_{1}/N_{1}<1/2$ or if ($n_{1}/N_{1}=1/2$ and $n_{2}/N_{2}<1/2).$
(Consequently, by symmetry, $\delta_{2}^{1}(n_{1},n_{2})=0$ if $n_{1}%
/N_{1}>1/2$ or ($n_{1}/N_{1}=1/2$ and $n_{2}/N_{2}>1/2)).$ For easy reference,
we refer to this initialization by SO (for \textquotedblleft
suboptimal\textquotedblright).\smallskip

In TABLE I, we report the minimax regret value using the formula in Stoye
(2009, Corollary 1) for the original model (before discretization) for the
subset $N_{1}=N_{2}\in\{5,60,100,200\}$ of the sample sizes we considered. In
addition, for each initialization (ES\ and SO) and each of the two weighting
schemes $\alpha_{n}$ we report the maximal regret of $\delta^{n}$ for
$n\in\{1,150,500,2000\}$ (over all actions by nature with $p=1000$)$.$ All
digits of maximal regret of $\delta^{n}$ that coincide with the ones of the
actual minimax regret value are displayed in bold, starting at the tenths
place, then the hundredths place, etc. until there is a
discrepancy.\footnote{The additional results for other $(N_{1},N_{2})$ values
appear in the Supplementary Appendix. As discussed in Comment 1 after
Algorithm \ref{specialized algorithm} minimax regret of the discretized model
is smaller than the minimax regret value (but converges to the latter as
$p\rightarrow\infty).$ That explains why in very few cases (like e.g. when
$N_{1}=N_{2}=5)$ the reported maximal regret is (very slightly, i.e. in the
10$^{-7}$ digit) smaller than the minimax regret value of the original model.}

The\textbf{ key findings} from the results in TABLE I are as follows. The
results reflect the convergence results from Proposition
\ref{proof of convergence}. As the number of iterations of the algorithm
increases the maximal regret of $\delta^{n}$ approaches the minimax regret
value. As suggested by the theory, this holds for both choices of $\alpha_{n}$
and initializations $\delta^{1}$ in all cases considered. We find that
$\alpha_{n}=(5+n)^{-.7}$ leads to faster convergence than $\alpha_{n}=1/n.$
Namely, maximal regret of $\delta^{500}$ and $\delta^{2000}$ is identical to
minimax regret up to at least four digits after the comma (and sometimes up to
seven digits after the comma) for the former weighting scheme. Quite
remarkably, maximal regret of $\delta^{150}$ already typically approximates
minimax regret up to four digits (in very few cases only up to three or even
up to five digits). Instead, for the latter weighting, maximal regret of
$\delta^{500}$ (and $\delta^{2000})$ coincides with minimax regret only up to
two or three (two to four) digits. Based on these results, we strongly suggest
using $\alpha_{n}=(5+n)^{-.7}$ over the original weighting $1/n$ proposed by
Robinson (1951). These results also reinforce an earlier remark about the
importance of using a good weighting scheme and potentially obtaining
theoretical insights on the convergence speed as a function of the weighting scheme.

Regarding the initialization, the values reported for maximal regret of
$\delta^{1}$ illustrate that SO is indeed a very suboptimal treatment rule.
E.g. for $(N_{1},N_{2})=60$ maximal regret under SO equals .3851399 while
under ES it equals .0170815. Somewhat surprisingly however, Algorithm
\ref{specialized algorithm} recovers quite quickly from the suboptimal
initialization and displays quite similar convergence patterns when started
with SO versus ES. When considering maximal regret of $\delta^{150}$ often the
algorithm started with SO leads to at least equal if not smaller values! A
reason for why the initialization may not matter that much is that no matter
what the initialization is, given the iterative structure of the algorithm,
the least favorable distribution of nature needs to build up from scratch. One
might want to consider modifications of the algorithm that also start off with
an informed guess on a least favorable distribution. Overall, the finding that
the initialization is not crucial for fast convergence is good news.

Substantial improvements in terms of maximal regret are obtained by the
algorithm after only quite few iterations. E.g. when $(N_{1},N_{2})=5$ and
$\alpha_{n}=(5+n)^{-.7},$ the maximal regret of ES\ equals .0703386 while
minimax regret equals .054308. After 150 iterations the algorithm generates a
treatment rule whose maximal regret equals .054309.

As mentioned already, $R(\delta^{n},\mu_{BR}^{n})$ is not monotonically
decreasing in $n.$ Likewise, $R^{n}$ is not monotonically increasing in $n$.
In practice therefore, after a certain number of iterations, $N$ say, one
should report the interval
\begin{equation}
I_{N}:=[\max_{n\leq N}R^{n},\min_{n\leq N}R(\delta^{n},\mu_{BR}^{n})]
\label{bound for minmax value}%
\end{equation}
that contains the minimax regret value (of the discretized game) and use the
treatment rule $\delta^{n}$ for the $n$ that minimizes $R(\delta^{n},\mu
_{BR}^{n})$ over all $n\leq N.$ For example, for ES and $\alpha_{n}%
=(5+n)^{-.7},$ $I_{2000}$ equals [.0542240,.0543086], [.0155017,.0155298],
[.0119668,.0120253], [.0083640,.0085001] for $(N_{1},N_{2})=5,$ $60,$ $100,$
$200,$ respectively. Note again here that the upper endpoint of $I_{N}$ maybe
smaller than the minimax regret value of the original (not discretized) model.
E.g. that happens for $(N_{1},N_{2})=200$ but only in the 10$^{-7}$ digit.

In all examples, not only this one, as one would expect, maximal regret
decreases in the sample size.

Regarding \textbf{computation time} (using $p=1000,$ ES,\ $\alpha
_{n}=(5+n)^{-.7},$ and 2000 iterations in all cases) for $(N_{1}%
,N_{2})=(10,10)$ it takes less than 5 minutes while for $(N_{1},N_{2}%
)=(200,200)$ it takes about 24 minutes. In the latter case, the precision for
maximal regret of $\delta^{2000}$ is about $3\mathbb{\times}10^{-6}$. This is
quite remarkable; namely, note that there are $2^{N_{1}(N_{2}+1)/2+N_{2}/2}$
symmetric treatment rules that take values $\delta_{2}(n_{1},n_{2})\in\{0,1\}$
except when $n_{1}/N_{1}=n_{2}/N_{2}=1/2$ (in which case $\delta_{2}%
(n_{1},n_{2})=1/2).$ With $(N_{1},N_{2})=(200,200)$ that number equals
$2^{20200}\approx10^{6080.8}!$

Note that even larger sample sizes are possible. E.g. we also considered
$(N_{1},N_{2})=(250,250)$, where maximal regret for $\delta^{n}$ equals
.008007, .007665, .007615, and .007603 for $n=1,150,500,$ and $2000,$
respectively, and $R(\delta^{n},\mu_{BR}^{n})-R^{n}$ equals .007668, .001310,
.000472, and .000223 for these choices of $n$ and computation time being about
39 minutes$.$ In that case $2^{N_{1}(N_{2}+1)/2+N_{2}/2}=2^{31500}%
\approx10^{9482.4}!$ The minimax regret value of the original model equals
.007602. When $N_{1}=N_{2}=300$ there are \ about $10^{13636.7}$ symmetric
\textquotedblleft nonrandomized\textquotedblright\ rules. The theoretical
minimax regret value equals .006940. Maximal regret of $\delta^{n}$ equals
.007279, .007053, .006950, and .006939 for $n=1,150,500,$ and $2000,$
respectively, and $R(\delta^{n},\mu_{BR}^{n})-R^{n}$ equals .006998, .001222,
.000352, and .000271 for these choices of $n$. Computation time is 51
minutes.\smallskip\ With $N_{1}=N_{2}=400$ computation time increases to 1
hour 27 minutes.

To give some idea about the complexity of least favorable distributions, when
$(N_{1},N_{2})=20,$ ES,\ and $\alpha_{n}=(5+n)^{-.7}$ the best response by
nature against $\delta^{2000\text{ }}$is a mixed strategy with $\overline
{m}=142$ in Proposition \ref{symmetry statement}.\bigskip

\textbf{TABLE\ I: }Maximal regret of $\delta^{n}$ for various choices of $n$,
$(N_{1},N_{2}),$ weighting schemes and initializations.%

\begin{tabular}
[c]{c||cccc}%
\ \ Weighting choice $\alpha_{n}$ \  & $\ \ \ \ \ n^{-1}$\ \ \ \ \  &
$\ \ \ \ \ n^{-1}$\ \ \ \ \  & $(5+n)^{-.7}$ & $(5+n)^{-.7}$\\
Initialization $\delta^{1}$ & SO & ES & SO & ES\\\hline\hline
\end{tabular}

$(N_{1},N_{2})=200;$\ minimax regret value=.00850086%

\begin{tabular}
[c]{c||cccc}\hline\hline
Maximal regret of $\delta^{1}$ & .4233904 & \textbf{.00}90009 & .4233904 &
\textbf{.00}90009\\
Maximal regret of $\delta^{150}$ & \textbf{.0}108846 & \textbf{.0}110614 &
\textbf{.0085}676 & \textbf{.008}6036\\
Maximal regret of $\delta^{500}$ & \textbf{.00}90704 & \textbf{.00}91066 &
\textbf{.0085}934 & \textbf{.00850}27\\
Maximal regret of $\delta^{2000}$ & \textbf{.008}6211 & \textbf{.008}6256 &
\textbf{.0085008} & \textbf{.00850}38\\\hline\hline
\end{tabular}

$(N_{1},N_{2})=100;$\ minimax regret value=.01202529%

\begin{tabular}
[c]{c||cccc}\hline\hline
Maximal regret of $\delta^{1}$ & .4022899 & \textbf{.012}9888 & .4022899 &
\textbf{.012}9888\\
Maximal regret of $\delta^{150}$ & \textbf{.01}31981 & \textbf{.01}44591 &
\textbf{.012}1011 & \textbf{.0120}440\\
Maximal regret of $\delta^{500}$ & \textbf{.012}3099 & \textbf{.012}5261 &
\textbf{.012025}8 & \textbf{.0120}691\\
Maximal regret of $\delta^{2000}$ & \textbf{.0120}876 & \textbf{.012}1344 &
\textbf{.012025}8 & \textbf{.01202}64\\\hline\hline
\end{tabular}

$(N_{1},N_{2})=60;$\ minimax regret value=.01553018%

\begin{tabular}
[c]{c||cccc}\hline\hline
Maximal regret of $\delta^{1}$ & .3851399 & \textbf{.01}70815 & .3851399 &
\textbf{.01}70815\\
Maximal regret of $\delta^{150}$ & \textbf{.01}62264 & \textbf{.01}72229 &
\textbf{.0155}947 & \textbf{.01553}77\\
Maximal regret of $\delta^{500}$ & \textbf{.015}6859 & \textbf{.015}9248 &
\textbf{.01553}18 & \textbf{.0155}883\\
Maximal regret of $\delta^{2000}$ & \textbf{.0155}676 & \textbf{.015}6287 &
\textbf{.0155301} & \textbf{.015530}6\\\hline\hline
\end{tabular}

$(N_{1},N_{2})=5;$\ minimax regret value=.05430889%

\begin{tabular}
[c]{c||cccc}\hline\hline
Maximal regret of $\delta^{1}$ & .2730320 & \textbf{.0}703386 & .2730320 &
\textbf{.0}703386\\
Maximal regret of $\delta^{150}$ & \textbf{.054}5794 & \textbf{.054}7494 &
.\textbf{0543}194 & \textbf{.0543}309\\
Maximal regret of $\delta^{500}$ & \textbf{.054}4255 & \textbf{.054}4402 &
.\textbf{0543}246 & .\textbf{0543}287\\
Maximal regret of $\delta^{2000}$ & \textbf{.0543}327 & \textbf{.0543}418 &
.\textbf{0543}110 & .\textbf{054308}7\\\hline\hline
\end{tabular}
\bigskip

Next, we apply Algorithm \ref{specialized algorithm} to \textbf{unbalanced
samples} for which no analytical formula is known for the minimax regret rule.
Namely, we consider all combinations of $N_{1}\in\{10,50,100\}$ and
$N_{2}-N_{1}\in\{10,50,100\}$ using the initialization as in Algorithm
\ref{specialized algorithm}. Based on the insights from the simulations in the
balanced case we use $\alpha_{n}=(5+n)^{-.7}.$ All reported results use
$p=1000$ again. Given we do not have an analytical formula for the minimax
regret value, TABLE II reports results on the maximal regret of $\delta^{n}$
together with the bound $R(\delta^{n},\mu_{BR}^{n})-R^{n}$ (for how far the
maximal regret of $\delta^{n}$ differs at most from the minimax regret value
of the discretized model) for $n\in\{1,150,500,2000\}$ and six of the nine
choices of $(N_{1},N_{2})$. The remaining results appear in the Supplementary
Appendix. In addition, we report $I_{2000}$ introduced in
(\ref{bound for minmax value}).\bigskip

\textbf{TABLE II: }Maximal regret of $\delta^{n}$ and $R(\delta^{n},\mu
_{BR}^{n})-R^{n}$ for several $n$ and $I_{2000}$ for several $(N_{1},N_{2}).$%

\begin{tabular}
[c]{c||ccc}%
$n$%
$\backslash$%
$(N_{1},N_{2})$ & $(10,20)$ & $(10,60)$ & $(10,110)$\\\hline\hline
$1$ & .037601;.034347 & .035049;.035049 & .035049;.035049\\
$150$ & .032765;.000074 & .028602;.001713 & .027716;.000452\\
$500$ & .032730;.000033 & .028580;.000003 & .027593;.000624\\
$2000$ & .032728;.000003 & .028580;.000003 & .027583;.000081\\
$I_{2000}$ & [.032726,.032727] & [.028578,.028579] &
[.027501,.027579]\\\hline\hline
\end{tabular}
\smallskip%

\begin{tabular}
[c]{cccc}\hline\hline
$n$%
$\backslash$%
$(N_{1},N_{2})$ & $(100,110)$ & $(100,150)$ & $(100,200)$\\\hline\hline
$1$ & .011793;.003706 & .011269;.010099 & .010921;.010379\\
$150$ & .011812;.000918 & .011112;.001011 & .010527;.001344\\
$500$ & .011766;.000104 & .010973;.000439 & .010404;.000422\\
$2000$ & .011739;.000058 & .010963;.000097 & .010403;.000144\\
$I_{2000}$ & [.011705,.011736] & [.010874,.010962] & [.010311,.010402]
\end{tabular}
\bigskip

The \textbf{key findings} from TABLE II are as follows. As expected from the
theory, maximal regret of $\delta^{n}$ converges (to the minimax regret value
of the discretized model) and $R(\delta^{n},\mu_{BR}^{n})-R^{n}$ goes to zero
as the number of iterations $n$ increases. In all cases considered
$R(\delta^{n},\mu_{BR}^{n})-R^{n}$ for $n=2000$ ranges between .000003 and
.000208. Furthermore, in all cases considered the width of $I_{2000}$ falls
into the interval $[10^{-6},$ $9\times10^{-5}].$ That implies that after 2000
iteration, one has found a treatment rule whose maximal regret is at most
$9\times10^{-5}$ larger than the minimax regret value of the (discretized) model.

We also considered Robinson's (1951) weighting, $\alpha_{n}=n^{-1}$.
Consistent with the findings above from balanced samples, in all cases
$\alpha_{n}=(5+n)^{-.7}$ produced smaller maximal regret for $\delta^{2000}.$
For brevity, we do not report those results.

To give some idea about the structure of the minimax regret rule, in the
Supplementary Appendix we report $\delta^{5000}$ for $N_{1}=10,$ $N_{2}=20$,
ES,\ and $\alpha_{n}=(5+n)^{-.7}$.

What are the gains of incorporating symmetry as in Algorithm
\ref{specialized algorithm} versus using Algorithm \ref{general algorithm} in
terms of computation time and maximal regret after a given number of
iterations? Consider e.g. the case $p=1000,$ ES, and $\alpha_{n}=(5+n)^{-.7}$.
When $(N_{1},N_{2})=(200,200)$ it takes 42 minutes to run 2000 simulations
(versus 24 minutes when symmetry is exploited) and maximal regret of
$\delta^{2000}$ equals .008505 (versus .008503 when symmetry is exploited).
For $(N_{1},N_{2})=(10,60)$ it takes 11 minutes to run 2000 simulations with
maximal regret of $\delta^{2000}$ equal to .028595 while the algorithm that
exploits symmetry only takes 5.5 minutes and maximal regret of $\delta^{2000}$
equals .028580. The advantage of Algorithm \ref{specialized algorithm} for
these sample sizes is therefore mostly in terms of computation
speed.\smallskip

Next, we provide some comparisons with the \textquotedblleft multiplicative
weights\textquotedblright\textbf{ }algorithm of Aradillas Fern\'{a}ndez,
Blanchet, Montiel Olea, Qiu, Stoye, and Tan (2024, ABMQST from now
on).\footnote{We would like to thank Pepe Montiel for sharing very helpful
insights related to this discussion.} Given the gigantic number of
nonrandomized treatment rules even for moderate numbers of $N_{1}$ and $N_{2}$
it is not possible to implement that algorithm when the policymaker is the
\textquotedblleft outer\textquotedblright\ player. However, using the identity
$\inf_{\delta\in\mathbb{D}}\sup_{s\in S}R(\delta,s)=\sup_{s\in S}\inf
_{\delta\in\mathbb{D}}R(\delta,s)$ discussed in the Appendix B of ABMQST, one
could let the policymaker play the role of the \textquotedblleft
inner\textquotedblright\ player and use comparisons of conditional means and
Bayes' rule to solve $\inf_{\delta\in\mathbb{D}}R(\delta,s)$ for given $s$ as
we suggest in (\ref{condition 0})-(\ref{condition}). Using that approach, the
subgradient and nature's strategy are $(p+1)^{T}$-dimensional vectors.
Calculating the former, requires $(p+1)^{T}$ many regret calculations (as does
grid search used in fictitious play). Storing the latter, while likely still
possible when $T=2$, is likely impossible when $T>2.$ On the other hand,
fictitious play only requires storing the strategies of the policymaker and
nature. The former can be characterized by a $(N_{1}+1)\times(N_{2}+1)$
matrix, while the latter (in our experiments) is very sparse with support only
on some 100 points after 2000 iterations. In contrast, by construction,
\textquotedblleft multiplicative weights\textquotedblright\textbf{\ }assigns
positive probability to \emph{each} of the $(p+1)^{T}$ actions by nature.
Thus, calculation of the sums appearing in (\ref{condition}) requires summing
$\overline{m}=(p+1)^{T}$ terms for each of the $(N_{1}+1)(N_{2}+1)$ arguments
of the treatment rule. Instead, using fictitious play, $\overline{m}$ (in our
experiments) is of the order of 100 even after 2000 iterations which leads to
considerably less computational effort.\footnote{Potentially, using some
thresholding approach could alleviate the computational burden for
\textquotedblleft multiplicative weights.\textquotedblright} Thirdly, note
that the maximal number of iterations $\lceil2\ln((p+1)^{T})/\epsilon
^{2}\rceil$ given in Theorem 1 of ABMQST that guarantees an $\epsilon$ minimax
regret rule equals $3.4\times10^{9}$ when $T=2,p=1000,$ and $\epsilon
=9\times10^{-5}$. Instead, recall that in all cases reported in TABLE\ II
after just 2000 iterations, a minimax regret rule is found whose maximal
regret is at most $9\times10^{-5}$ larger than the minimax regret value of the
(discretized) model. Conversely, if for $T=2$ and $p=1000$ one sets
$\lceil2\ln((p+1)^{T})/\epsilon^{2}\rceil$ equal to 2000 one obtains
$\epsilon=0.118.$

In general, we believe that a comparison of two algorithms and the notion of
optimality of an algorithm, should take into account the number of iterations
\emph{and} the number of operations performed in each iteration. The upper
bound $\lceil2\ln((p+1)^{T})/\epsilon^{2}\rceil$ given in Theorem 1 of ABMQST
is a powerful result that applies to a broad class of models. But it does not
seem to be binding in the specific application considered here.

\subsubsection{Restricted strategy space for nature \label{restrict}}

We now consider the case where the set $M$ of mean vectors no longer equals
the unrestricted set $[0,1]^{2}$. Instead, we examine a scenario where the
policymaker uses a priori information about certain parametric restriction on
$(\mu_{1},\mu_{2}).$ Being able to accommodate such restrictions is highly
empirically relevant. For example, average wage after job training should be
expected to be nonsmaller. If one also factors in a fixed cost for job
training into the analysis one ends up with a set of particular restrictions
on the mean vectors $(\mu_{1},\mu_{2})$. For the purpose of illustration, we
take
\begin{equation}
M=\{(\mu_{1},\mu_{2})\in\lbrack0,1]^{2}:0.9\mu_{1}\leq\mu_{2}\leq1.2\mu_{1}\}.
\label{restriction M}%
\end{equation}
For restricted $M$, Proposition \ref{symmetry statement} that relied heavily
on symmetry, no longer applies. Therefore, we do not restrict the search to
symmetric rules and use Algorithm \ref{general algorithm} to approximate the
minimax regret rule.

We consider sample sizes $N_{1}=N_{2}\in\{5,10,20,50\},$ $(N_{1},N_{2}%
)\in\{(5,10),(10,5),(10,50),(50,10)\},$ $\alpha_{n}=(5+n)^{-.7},$ $p=1000,$
and use 2,000 iterations. We initialize with $\delta^{1}$ equal to the
ES\ rule defined by $\delta_{2}^{1}(n_{1},n_{2})=1$ if $n_{1}/N_{1}\leq
n_{2}/N_{2}$ and $\delta_{2}^{1}(n_{1},n_{2})=0$ otherwise and take $D$ to be
the set of all nonrandomized rules. We take
\begin{equation}
M_{\varepsilon}=\{(\mu_{1},\mu_{2})\in M;\mu_{j}=p_{j}/p\text{ for some }%
p_{j}\in\{0,1,...,p\}\text{ for }j=1,2\}.
\end{equation}
TABLE\ III gives results on the maximal regret of $\delta^{n}$ and the bound
$R(\delta^{n},\mu_{BR}^{n})-R^{n}$ for $n\in\{1,150,500,2000\}.$\bigskip

\textbf{TABLE III: }Maximal regret of $\delta^{n}$ and $R(\delta^{n},\mu
_{BR}^{n})-R^{n}$ for $n\in\{1,150,500,2000\}.$%

\begin{tabular}
[c]{c||cccc}\hline\hline
$n$%
$\backslash$%
$(N_{1},N_{2})$ & $(5,5)$ & $(10,10)$ & $(20,20)$ & $(50,50)$\\\hline\hline
$1$ & .059049;.059049 & .036090;.036090 & .027256;.027256 & .018518;.018518\\
$150$ & .036396;.000623 & .030029;.000665 & .023947;.000492 &
.016758;.000809\\
$500$ & .036159;.000275 & .029816;.000541 & .023775;.000182 &
.016601;.000743\\
$2000$ & .035986;.000173 & .029789;.000100 & .023788;.000119 & .016595;.000131
\end{tabular}
\medskip%

\begin{tabular}
[c]{c||cccc}\hline\hline
$n$%
$\backslash$%
$(N_{1},N_{2})$ & $(5,10)$ & $(10,5)$ & $(10,50)$ & $(50,10)$\\\hline\hline
$1$ & .035049;.035049 & .059049;.059049 & .027084;.027084 & .035049;.035049\\
$150$ & .033472;.000542 & .032834;.001307 & .025716;.000195 &
.024615;.000683\\
$500$ & .033415;.000170 & .032579;.000217 & .025690;.000342 &
.024567;.000422\\
$2000$ & .033388;.000153 & .032552;.000147 & .025574;.000116 & .024529;.000195
\end{tabular}
\bigskip

As the \textbf{key findings} TABLE\ III\ illustrates again the convergence
properties of maximal regret of $\delta^{n}$ and convergence to zero of
$R(\delta^{n},\mu_{BR}^{n})-R^{n}$ as $n$ increases. After 2000 iterations the
algorithm has reached treatment rules $\delta^{2000}$ whose maximal regret is
at most $1.95\mathbb{\times}(10^{-4})$ away from the minimax regret value$.$
Compared to TABLE\ I for equal sample sizes we find that here maximal regret
of $\delta^{n}$ is always smaller given that nature is less powerful here
given the restrictions on its parameter space.

Furthermore, the results from TABLE III illustrate that when there is a priori
information the ES rule may be a very suboptimal choice. For example when
$(N_{1},N_{2})=(5,5)$ or $(N_{1},N_{2})=(10,5)$ its maximal regret equals
.059049, more than 1.6 times or 1.8 times as much as the maximal regret of
$\delta^{2000}$ which equals .035986 or .032552. With $M$ being restricted it
is very hard to guess a reasonable treatment rule. This provides strong
motivation for the algorithm provided in this paper whose convergence
properties do not seem to be affected much by the initialization.

We describe $\delta_{2}^{2000}$ in the case where $N_{1}=N_{2}=5.$ We find
that $\delta_{2}^{2000}(n_{1},n_{2})=1$ if $n_{1}<n_{2}$ and $\delta
_{2}^{2000}(n_{1},n_{2})=0$ if $n_{1}>n_{2}.$ Finally, $\delta_{2}%
^{2000}(n_{1},n_{1})$ equals 1, .948767, .906576, .844678, .704494, and
.533296 when $n_{1}=0,1,...,5,$ respectively. Next we (partially) describe
$\delta_{2}^{2000}$ in the case where $N_{1}=5$ and $N_{2}=10.$ In that case,
$n_{1}/N_{1}<n_{2}/N_{2}$ still implies $\delta_{2}^{2000}(n_{1},n_{2})=1$ but
it is no longer true that for all 30 cases where $n_{1}/N_{1}>n_{2}/N_{2},$
always $\delta_{2}^{2000}(n_{1},n_{2})=0$ holds$.$ In five of those cases
$\delta_{2}^{2000}(n_{1},n_{2})>0.$\footnote{We ran the algorithm for 4000
iterations and now only two such cases (up to 10$^{-6}$ precision) occured.
Namely $\delta_{2}^{4000}(1,0)=1$ and $\delta_{2}^{4000}(1,1)=.000280.$ So,
when 1 success for treatment 1 and zero successes for treatment 2 are
observed, treatment 2 should be chosen with probability 1 - that seems quite
counterintuitive.}

Even though we cannot exploit any symmetry conditions for this scenario, the
algorithm is still running very quickly.\footnote{What significantly reduces
the running time is that the grid search over mean vectors when updating
nature's strategy requires less time given the imposed restrictions on $M$.}
E.g. it takes less than 4 and 4.5 minutes to obtain 2000 iterations when
$N_{1}=N_{2}=5$ and $50$, respectively; virtually the same computation time in
both cases.

\section{Treatment assignment when testing
innovations\label{testing two innovations}}

Suppose the policymaker knows the mean outcome $\mu_{T}$ of a status quo
treatment. She needs to consider whether to switch from the status quo to one
of $T-1$ alternative treatments whose mean outcomes are unknown. Stoye (2009)
considers the case $T=2$ and works out a minimax regret rule. In this section,
we apply Algorithm \ref{general algorithm} to $T>2$ where a sample of equal
sizes on $T-1$ alternative treatments is observed. Potential outcomes are
elements of $\{0,1\};$ the general case where outcomes are in $[0,1]$ is then
dealt with via the coarsening approach of Proposition \ref{coarsening}. A
priori information on restrictions of the mean vector $\{\mu_{1},...,\mu
_{T-1}\}$ can be incorporated. Without further restrictions $M=[0,1]^{T-1}%
\times\mu_{T}.$ As in the previous example, to reduce the computational
effort, we derive certain symmetry conditions first.

As before, we can identify the DGP by the vector of means. The observed sample
$w_{N}$ can be summarized by the number of observed successes $n_{t}$ for
treatments $t=1,...,T-1$. For a rule $\delta$ and sample $w_{N}$ we denote by
$\delta_{t}(w_{N})$ the probability that treatment $t$ (for $t=1,...,T-1)$
will be chosen and consequently by $1-\sum_{t=1}^{T-1}\delta_{t}(w_{N})$ the
probability that the status quo persists. We suppress the dependence on
$\mu_{T}$ in the notation when deriving a minimax regret rule for a particular
$\mu_{T}.$

\subsection{Symmetry restrictions}

There are $T$ treatments to choose from and treatment $T$ is the known status
quo$.$ Data of equal sample size $\overline{N}$ is observed for each of the
$T-1$ innovations. We take $S=\{0,1\}^{T-1}$. Denote by $\sigma^{-1}$ the
inverse of a permutation $\sigma$ in the group of permutations. We make the
following definition.\smallskip

\textbf{Definition (symmetric treatment rule).} \emph{We call a treatment rule
}$\delta$\emph{ symmetric\footnote{It should not be confusing that the notion
of symmetry changes in different applications.} if whenever the vectors
}$w_{N}=(n_{1},...,n_{T-1})$\emph{ and }$w_{N}^{\prime}=(n_{1}^{\prime
},...,n_{T-1}^{\prime})$\emph{ are permutations of each other, i.e.
}$n_{\sigma(t)}=n_{t}^{\prime}$\emph{ for some permutation }$\sigma$\emph{ on
}$\{1,...,T-1\}$ \emph{then}
\begin{equation}
\delta_{t}(w_{N})=\delta_{\sigma^{-1}(t)}(w_{N}^{\prime})\text{ for }%
t\in\{1,...,T-1\}. \label{symmetry def in}%
\end{equation}

\textbf{Comments.} 1. As in Section \ref{lead example}, the symmetry property
allows for computational gains in approximating minimax regret rules. Note
that (\ref{symmetry def in}) implies that $\delta_{T}(w_{N})=\delta_{T}%
(w_{N}^{\prime}).$ In the case $T=3$, symmetry requires that $\delta
_{1}((n_{1},n_{2}))=\delta_{2}((n_{2},n_{1}))$ for all $n_{1},n_{2}%
\in\{0,1,...,\overline{N}\}.$ Thus, for a symmetric treatment rule $\delta,$
given values $\delta_{2}(w_{N})$ and $\delta_{3}(w_{N})$ for $w_{N}%
=(n_{1},n_{2})$ one can determine $\delta_{1}(w_{N})$ and $\delta_{t}%
((n_{2},n_{1}))$ for $t\in\{1,2,3\}.$ Therefore, knowing $\delta_{t}%
((n_{1},n_{2}))$ for $t\in\{2,3\}$ and all $n_{2}\geq n_{1}$ one automatically
knows the entire treatment rule.\medskip

The next proposition establishes that one can find an action pair $(\delta,s)$
with $\delta\in\mathbb{D}$ and $s\in\Delta\mathbb{S}$ for the policymaker and
nature, respectively, that constitutes a Nash equilibrium and where $\delta$
is symmetric according to the definition just given.

\begin{proposition}
\label{symmetry statement in}$($i$)$ Suppose nature plays a mixed strategy
that picks $m$ mixed strategies $\mu_{m}^{MS}$ with probabilities $p_{m}>0$
for $m=1,...,\overline{m}$ for some $\overline{m}\geq1,$ where $p_{1}%
+...+p_{\overline{m}}=1$ and for all $m=1,...,\overline{m},$ $\mu_{m}^{MS}$
mixes uniformly over all of the $(T-1)!$ permutations $(\mu_{\sigma
(1)m},...,\mu_{\sigma(T-1)m},\mu_{T})$ of a vector $(\mu_{1m},...,\mu
_{T-1m},\mu_{T}),$ where $\sigma$ denotes a permutation on $\{1,...,T-1\}.$
Then if there exists a treatment rule $\delta$ that is a best response to that
strategy $($defined by the $\mu_{m}^{MS}$ and $p_{m}$ for $m=1,...,\overline
{m})$ then it can be chosen to satisfy $($\ref{symmetry def in}$)$.

$($ii$)$ Assume the policymaker chooses a symmetric treatment rule $\delta$
and $(\mu_{1},...,\mu_{T})$ is a best response by nature. Then for any
permutation $\sigma$ on $\{1,...,T-1\}$ the mean vector $(\mu_{\sigma
(1)},...,\mu_{\sigma(T-1)},\mu_{T})$ is also a best response.
\end{proposition}

\textbf{Comments. }1. Proposition \ref{symmetry statement in} establishes that
a Nash equilibrium of the unrestricted game exists in which the policymaker
plays a symmetric treatment rule. Therefore, if there exists a minimax regret
rule that is part of a Nash equilibrium then
\begin{equation}
\inf_{\delta\in\mathbb{D}}\max_{\mu\in\lbrack0,1]^{T-1}\times\mu_{T}}%
R(\delta,\mu)=\inf_{\substack{\delta\in\mathbb{D}\\\delta\text{ is symmetric}%
}}\max_{\substack{s\in\Delta\mathbb{S}\\s\text{ is as in Proposition
\ref{symmetry statement in}(i)}}}R(\delta,\mu). \label{reformulation}%
\end{equation}
Using the formulation of the problem as on the right side of the equation in
(\ref{reformulation}), one can pick as $D$ the finite set of symmetric
nonrandomized rules. The proof of part (ii) follows from the fact that if
$\delta$ is symmetric then $R(\delta,(\mu_{1},...,\mu_{T}))=R(\delta
,(\mu_{\sigma(1)},...,\mu_{\sigma(T-1)},\mu_{T}))$ for any permutation
$\sigma$ on $\{1,...,T-1\}.$

\subsection{Algorithm}

We establish that the assumption of Lemma \ref{lem:finite approx} about the
regret function $R(\delta,\cdot)$ being continuous in $\mu\in\lbrack
0,1]^{T-1}\times\mu_{T}$ uniformly in $\delta\in\mathbb{D}$ is
satisfied.\smallskip

\begin{lemma}
\label{lem:continuity in innovation}$\forall\lambda>0$ there exists $\eta>0$
such that when $\Vert\mu-\mu^{\prime}\Vert<\eta$ for $\mu,\mu^{\prime}%
\in\lbrack0,1]^{T-1}\times\mu_{T}$ then for all $\delta\in\mathbb{D}$ we have
$|R(\delta,\mu)-R(\delta,\mu^{\prime})|<\lambda$.\medskip
\end{lemma}

We will employ the analogous discretization strategy given in
(\ref{eps discretization}). For given $\mu_{T}\in\lbrack0,1],$ some
$p\in\mathbb{N}$ and $\varepsilon=1/p$ define%
\begin{equation}
M_{\varepsilon}=\{(\mu_{1},...,\mu_{T-1},\mu_{T});\mu_{t}=p_{t}/p\text{ for
some }p_{t}\in\{0,1,...,p\}\text{ and }t=1,...,T-1\}.
\label{M eps second example}%
\end{equation}
We modify the general algorithm to reflect the symmetry property under the
setup of testing $T-1$ innovations. Pick again a tolerance level $\xi>0.$

\begin{algorithm}
\label{specialized algorithm in}

\textbf{Initialization:}

i$)$ Define $\delta^{1}$ as the empirical success rule. More precisely, let
$\overline{n}=\max\{n_{1}/\overline{N},...,n_{T-1}/\overline{N},\mu_{T}\}$ and
define $W$ by letting $t\in W\subset\{1,...,T\}$ if $n_{t}/\overline
{N}=\overline{n}$ for $t\in\{1,...,T-1\}$ and $\mu_{T}=\overline{n}$ for
$t=T.$ Then
\begin{equation}
\delta_{t}^{1}(w_{N})=1/|W|\text{ if }t\in W\text{ and }\delta_{t}^{1}%
(w_{N})=0\text{ otherwise.}%
\end{equation}

ii$)$ Initialize $\nu^{0}=0$ and $R^{0}=0.$

\textbf{Iteration: }

\textbf{For }$n=1,2,3...$ \textbf{DO}:\textbf{ }

i$)$ Find a best response $\mu_{BR}^{n}$ by nature in response to $\delta
^{n},$ where $\mu_{BR}^{n}$ equals a mixed strategy that randomizes uniformly
over all $(T-1)!$ permutations $(\mu_{\sigma(1)}^{n},...,\mu_{\sigma(T-1)}%
^{n},\mu_{T})$ of the first $T-1$ components of some vector $(\mu_{1}%
^{n},...,\mu_{T-1}^{n},\mu_{T})$ in $M_{\varepsilon}$.

ii$)$ If
\begin{equation}
R(\delta^{n},\mu_{BR}^{n})-R^{n-1}<\xi
\end{equation}
then \textbf{break. }Use the rule $\delta^{n}.$

iii$)$ Update nature's mixed strategy to be the following weighted average
\begin{equation}
\nu^{n}=(1-\alpha_{n})\nu^{n-1}+(\alpha_{n}/(T-1)!)%
{\textstyle\sum\nolimits_{\sigma}}
I((\mu_{\sigma(1)}^{n},...,\mu_{\sigma(T-1)}^{n},\mu_{T}^{n})),
\end{equation}
where $I(x)$ denotes a point mass of size 1 at the point $x\in\lbrack0,1]^{T}$
and the sum is over all permutations of $(1,...,T-1)$.

iv$)$ Compute a best response $\delta_{BR}^{n}$ in nonrandomized symmetric
strategies by the player to $\nu^{n}$, that is,
\begin{equation}
R^{n}:=R(\delta_{BR}^{n},\nu^{n})=\min_{_{\substack{\delta\in\mathbb{D}%
,\delta\text{ is nonrandomized}\\\text{and symmetric}}}}R(\delta,\nu^{n}).
\end{equation}
v$)$ Update the treatment rule by%
\begin{equation}
\delta^{n+1}=(1-\alpha_{n+1})\delta^{n}+\alpha_{n+1}\delta_{BR}^{n}.
\end{equation}

\end{algorithm}

\textbf{Comments.} 1. In each iteration, the best response of the policymaker
can be calculated via Bayes rule using (\ref{bayes in example}). The best
response by nature is solved again by grid search over all $(p+1)^{T-1}$
possible choices of mean vectors.

2. If $M$ is a strict subset of $[0,1]^{T-1}\times\mu_{T}$ the symmetry
property can in general not be exploited and Algorithm \ref{general algorithm}
is used instead.

3. Note that the existence of best responses of the particular type in
iteration steps i) and iv) is guaranteed by Proposition
\ref{symmetry statement in}.\medskip

We implement Algorithm \ref{specialized algorithm in} for the case $T=3$,
sample sizes $\overline{N}=N_{1}=N_{2}\in\{5,10,20,30,$ $40,50,100,200\}$ and
known mean of the status quo treatment equal to $\mu_{3}\in\{.2,.5,.8\}.$ The
initialization rule is the empirical success rule described in the beginning
of Algorithm \ref{specialized algorithm in}. We choose $p=1000$ and
$\alpha_{n}=(5+n)^{-.7}$. TABLE IV reports\textbf{ }maximal regret of
$\delta^{n}$ and $R(\delta^{n},\mu_{BR}^{n})-R^{n}$ for various choices of $n$
and $I_{2000}$ for a subset of the sample sizes.\footnote{To save space we
report the remaining results in the Supplementary Appendix.}$\bigskip$

\textbf{TABLE IV: }Maximal regret of $\delta^{n}$ and $R(\delta^{n},\mu
_{BR}^{n})-R^{n}$ for several $n$ and $I_{2000}$ for several $\overline{N}$
and $\mu_{3}.$%

\begin{tabular}
[c]{c||ccc}\hline\hline
$\quad\ \mu_{3}\qquad$ & \quad\ \ \ \ \ \ \ \ 0.2 \ \ \ \ \ \quad &
\quad\ \ \ \ \ \ \ \ \ 0.5 \ \ \ \ \ \ \quad & \quad\ \ \ \ \ \ \ \ \ 0.8
\ \ \ \ \ \ \quad
\end{tabular}

\begin{tabular}
[c]{c||ccc}\hline\hline
\multicolumn{4}{l}{$\overline{N}=200$}\\\hline\hline
$n=1$ & .008615;.001698 & .010715;.010715 & .0090221;.009022\\
$n=150$ & .008500;.000046 & .011725;.002386 & .008726;.001326\\
$n=500$ & .008500;.000042 & .010495;.000564 & .008397;.000888\\
$n=2000$ & .008500;.000043 & .010492;.000159 & .008374;.000171\\
$I_{2000}$ & [.008499;.008500] & [.010417;.010488] & [.008241;.008372]
\end{tabular}

\begin{tabular}
[c]{c||ccc}\hline\hline
\multicolumn{4}{l}{$\overline{N}=100$}\\\hline\hline
$n=1$ & .012333;.002467 & .015157;.015157 & .013036;.013036\\
$n=150$ & .012025;.000046 & .014872;.000681 & .011889;.000784\\
$n=500$ & .012025;.000000 & .014860;.000252 & .011841;.000180\\
$n=2000$ & .012025;.000002 & .014831;.000081 & .011839;.000166\\
$I_{2000}$ & [.012025;.012025] & [.014771;.014830] & [.011822.011835]
\end{tabular}

\begin{tabular}
[c]{c||ccc}\hline\hline
\multicolumn{4}{l}{$\overline{N}=50$}\\\hline\hline
$n=1$ & .017768;.003592 & .021440;.021440 & .019001;.019001\\
$n=150$ & .017057;.000111 & .021054;.000321 & .016804;.000155\\
$n=500$ & .017015;.000040 & .020978;.000045 & .016775;.000448\\
$n=2000$ & .017015;.000016 & .020967;.000073 & .016736;.000166\\
$I_{2000}$ & [.017010;.017015] & [.020963;.020966] & [.016725;.016729]
\end{tabular}

\begin{tabular}
[c]{c||ccc}\hline\hline
\multicolumn{4}{l}{$\overline{N}=10$}\\\hline\hline
$n=1$ & .043262;.009139 & .047978;.047978 & .047926;.047926\\
$n=150$ & .038815;.000303 & .047061;.000290 & .037808;.000808\\
$n=500$ & .038832;.000110 & .047365;.000498 & .037655;.000151\\
$n=2000$ & .038819;.000046 & .046945;.000132 & .037631;.000109\\
$I_{2000}$ & [.038810;.038810] & [.046936;.046936] & [.037537;.037549]
\end{tabular}
$\bigskip$

As the \textbf{key findings} TABLE\ IV\ illustrates again the convergence
properties of maximal regret of $\delta^{n}$ and convergence to zero of
$R(\delta^{n},\mu_{BR}^{n})-R^{n}$ as $n$ increases. After 2000 iterations the
algorithm has reached treatment rules $\delta^{2000}$ whose maximal regret is
at most $1.7\mathbb{\times}(10^{-4})$ away from the minimax regret value.
TABLE\ IV\ also illustrates that it is beneficial to keep track of the maximal
regret of the treatment rules along all iterations $n$ as the right boundary
of $I_{2000}$ maybe strictly smaller than maximal regret of $\delta^{2000}$
indicating that a treatment rule with smaller maximal regret occurred in an
earlier iteration. What is the impact of $\mu_{3}$ on the results? The minimax
regret value when $\mu_{3}=0.5$ is the largest (amongst those considered)
followed by the minimax regret value when $\mu_{2}=0.2$ and $\mu_{1}=0.8$,
although the difference is very small among the latter two.

\section{Conclusion}

We propose an algorithm to numerically approximate minimax regret rules and
prove that the maximal regret of the treatment rules generated by the
algorithm converges to the minimax regret of the discretized model. The latter
converges to the minimax regret value of the original model as the
discretization becomes finer. We illustrate in several examples that the
algorithm works very well in practice. Importantly, our framework allows
incorporation of a priori information that the policymaker can use regarding
the set of DGPs that nature can choose from. Also, the framework allows for a
restricted set of policy rules that the policymaker can choose from,
reflecting the important case where policy restrictions prevent the
policymaker from using certain rules. However, in the latter case one would in
general not be able to use the approach based on Bayes' rule.

There are several important open questions that go beyond the scope of the
current paper and require future research. In particular, we found that the
choice of weights $\alpha_{n}$ in the updating step of the algorithm
significantly impacted the convergence speed. Theoretical results should be
developed to work out optimal choices for the weights. Second, we found that
certain symmetry conditions can be exploited to significantly reduce the
complexity of the algorithm. Techniques need to be developed to detect all
such symmetries. Third, implementation details of the algorithm under policy
restrictions are quite important in cases where the approach via Bayes' rule
is no longer applicable. Fourth, progress has to be made in allowing for
covariates in the model.

There are many additional examples and related applications that the algorithm
or variants of it could be applied to. In particular, the proposed algorithm
may be helpful in approximating optimal tests in terms of weighted average
power when the null hypothesis is composite.

\section{Appendix}

To simplify the presentation, we typically do not index expectations and
probabilities by the DGP $s$ unless it is needed for clarity.\medskip

\textbf{Proof of Lemma \ref{lem:finite approx}.} First consider part (i).
Clearly, $V\geq V_{m}.$ Pick a $\lambda>0.$ We need to show that there exist
an $m_{\lambda}\in\mathbb{N}$ such that for all $m\geq m_{\lambda}$ we have
$V-V_{m}\leq\lambda.$ For every $\delta\in\mathbb{D}$ let $\mu_{\delta}\in M$
be such that $\max_{\mu\in M}R(\delta,\mu)=R(\delta,\mu_{\delta}).$ By the
assumed continuity of $R(\delta,\cdot)$ there exists $\eta_{\lambda}>0$ such
that $|R(\delta,\mu)-R(\delta,\mu^{\prime})|<\lambda$ whenever $\Vert\mu
-\mu^{\prime}\Vert<\eta_{\lambda}$. Pick $m_{\lambda}\in\mathbb{N}$ such that
for all $m\geq m_{\lambda},$ $\varepsilon_{m}<\eta_{\lambda}.$ Because
$M_{\varepsilon}$ is an $\varepsilon$-discretization of $M$ we can find
$\mu_{\delta,\varepsilon_{m}}\in M_{\varepsilon_{m}}$ such that $\Vert
\mu_{\delta}-\mu_{\delta,\varepsilon_{m}}\Vert<\varepsilon_{m}<\eta_{\lambda}$
and therefore%
\begin{equation}
\max_{\mu\in M_{\varepsilon_{m}}}R(\delta,\mu)\geq R(\delta,\mu_{\delta
,\varepsilon_{m}})\geq R(\delta,\mu_{\delta})-\lambda=\max_{\mu\in M}%
R(\delta,\mu)-\lambda. \label{step1}%
\end{equation}
That implies that $V_{m}\geq V-\lambda$ as desired. For part (ii) let
$\delta_{m}$ be any rule that satisfies $\max_{\mu\in M}R(\delta_{m}%
,\mu)\rightarrow V$ as $m\rightarrow\infty$. Note that for all $m\geq
\overline{m}_{\lambda}$ for some $\overline{m}_{\lambda}\in\mathbb{N}$
\begin{equation}
\max_{\mu\in M}R(\delta_{\varepsilon_{m}},\mu)\leq\max_{\mu\in M_{\varepsilon
_{m}}}R(\delta_{\varepsilon_{m}},\mu)+\lambda\leq\max_{\mu\in M_{\varepsilon
_{m}}}R(\delta_{m},\mu)+2\lambda, \label{step2}%
\end{equation}
where the first inequality holds by (\ref{step1}) for all $m\geq m_{\lambda}$
with $\delta_{\varepsilon_{m}}$ playing the role of $\delta.$ The second
inequality in (\ref{step2}) holds because by definition $\delta_{\varepsilon
_{m}}$ is approximately minimax regret for the case where the parameter space
for nature equals $M_{\varepsilon_{m}}$. It follows that $\max_{\mu\in
M}R(\delta_{\varepsilon_{m}},\mu)$ $\leq\max_{\mu\in M}R(\delta_{m}%
,\mu)+2\lambda$ for all $m\geq\overline{m}_{\lambda}$ which implies the
desired result. $\square$\medskip

\textbf{Proof of Proposition \ref{coarsening}. }We first consider the case of
fixed assignment. \textbf{i) }The distribution of $\widetilde{Y}_{t_{i},i}$ is
Bernoulli and by the law of iterated expectations the success probability
equals $\mu_{t_{i}}.$ Therefore, the distribution of the sample $\widetilde{w}%
_{N}$ depends on the state of nature $s$ only via $(\mu_{1},...,\mu_{T}).$
Denote by $\widetilde{\delta}$ the \textquotedblleft coarsened
version\textquotedblright\ of $\delta$ for an arbitrary $\delta\in\mathbb{D}$.
Given
\begin{equation}
R(\widetilde{\delta},s)=\max_{t\in\mathbb{T}}\{\mu_{t}\}-%
{\textstyle\sum\nolimits_{t=1}^{T}}
\mu_{t}E\delta_{t}(\widetilde{w}_{N}), \label{R of deltatilda}%
\end{equation}
also $R(\widetilde{\delta},s)$ depends on $s$ only via $\mu=(\mu_{1}%
,...,\mu_{T}).$ Denote by $s^{\prime}$ the distribution obtained from
$s\in\mathbb{S}$ by replacing each marginal $Y_{t}$ for $t\in\mathbb{T}$ with
a Bernoulli distribution with the same expectation and let $\mathbb{S}%
^{\prime}$ be the space of all such distributions $s^{\prime}$. Given
$\mathbb{S}$ is unrestricted (except for $\mu\in M)$, it follows that
$\mathbb{S}^{\prime}\subset\mathbb{S}.$ Because $s$ and $s^{\prime}$ have the
same mean vector, $R(\widetilde{\delta},s)=R(\widetilde{\delta},s^{\prime})$.
Because under $s^{\prime}$ it follows that $\widetilde{w}_{N}=w_{N}$ with
probability 1 we obtain $R(\widetilde{\delta},s^{\prime})=R(\delta,s^{\prime
})$ and thus
\begin{equation}
R(\widetilde{\delta},s)=R(\delta,s^{\prime}). \label{important equality}%
\end{equation}
Clearly,
\begin{equation}
\min_{\delta\in\mathbb{D}}\max_{s^{\prime}\in\mathbb{S}^{\prime}}%
R(\delta,s^{\prime})\leq\min_{\delta\in\mathbb{D}}\max_{s\in\mathbb{S}%
}R(\delta,s)\leq\min_{\widetilde{\delta}:\delta\in\mathbb{D}}\max
_{s\in\mathbb{S}}R(\widetilde{\delta},s). \label{string of inequal}%
\end{equation}
But because of $R(\widetilde{\delta},s)=R(\delta,s^{\prime})$ the inequalities
in (\ref{string of inequal}) actually hold as equalities and $\delta\in
\arg\min_{\delta\in\mathbb{D}}\max_{s^{\prime}\in\mathbb{S}^{\prime}}%
R(\delta,s^{\prime})$ implies $\widetilde{\delta}\in\arg\min_{\delta
\in\mathbb{D}}\max_{s\in\mathbb{S}}R(\delta,s)$. Therefore, $\delta^{C}$, the
coarsened version of a minimax regret rule in the binomial case, is minimax
regret when outcomes live on the unit interval.

\textbf{ii)} As in (\ref{important equality}),$\ R(\delta_{\varepsilon}%
^{C},s)=R(\delta_{\varepsilon},s^{\prime}),$ where $s^{\prime}$ is again the
distribution obtained from $s\in\mathbb{S}$ by replacing each marginal $Y_{t}$
with a Bernoulli distribution with the same expectation. Therefore,
\begin{equation}
\max_{s\in\mathbb{S}}R(\delta_{\varepsilon}^{C},s)=\max_{s^{\prime}%
,s\in\mathbb{S}}R(\delta_{\varepsilon},s^{\prime})\leq\min_{\delta
\in\mathbb{D}}\max_{s^{\prime}\in\mathbb{S}^{\prime}}R(\delta,s^{\prime
})+\varepsilon=\min_{\delta\in\mathbb{D}}\max_{s\in\mathbb{S}}R(\delta
,s)+\varepsilon, \label{string of inequali}%
\end{equation}
where the inequality holds by assumption on $\delta_{\varepsilon}$ and the
second equality follows because (\ref{string of inequal}) holds with
equalities. \smallskip

Next, we consider the case of random assignment, $P(t_{i}=t)=p_{t}$ for all
$i=1,...,N$ and $t\in\mathbb{T}.$ The proof of i) and ii) is identical to the
one just given. The distribution of $\widetilde{Y}_{t_{i},i}$ is Bernoulli and
by the law of iterated expectations the success probability equals $%
{\textstyle\sum\nolimits_{t\in\mathbb{T}}}
p_{t}\mu_{t}$ which equals the expectation of $Y_{t_{i},i}$ when we consider
randomness both in treatment assignment and outcomes. In particular, again,
given the particular sampling scheme, the distribution of the sample
$\widetilde{w}_{N}$ depends on the state of nature $s$ only via $(\mu
_{1},...,\mu_{T}).$ The proof for testing innovations is also identical.
$\square$\medskip

\textbf{Proof of Proposition \ref{analytic result}. }Assume nature mixes
evenly between the two vectors of means $(0,1/2)$ and $(1,1/2)$ and the
decision maker employs $\delta_{2}$ as described in Proposition
\ref{analytic result}. It is enough to establish that the two actions by
nature and the policymaker are best responses to each other. Note that
$\delta$ as defined is a symmetric treatment rule as defined in
(\ref{symmetry def}) and therefore Proposition \ref{symmetry statement}(ii) applies.

We now show that the vector of means $(0,1/2)$ maximizes regret given the
treatment rule $\delta$. Without loss of generality, we focus on the case
where $\mu_{1}<\mu_{2}$. Note that as below (\ref{formula regret})
\begin{equation}
R(\delta,(\mu_{1},\mu_{2}))=(\mu_{2}-\mu_{1})\sum_{n_{2}=0}^{N_{2}}%
(1-\delta_{2}(0,n_{2}))\binom{N_{2}}{n_{2}}\mu_{2}^{n_{2}}(1-\mu_{2}%
)^{N_{2}-n_{2}}. \label{form regret}%
\end{equation}
In particular, $R(\delta,(\mu_{1},\mu_{2}))$ is a linear and strictly
decreasing function of $\mu_{1}.$ Therefore, in order to maximize regret,
nature must choose $\mu_{1}=0$. Plugging in $\mu_{1}=0$ and $\delta
_{2}(0,n_{2})=n_{2}/N_{2},$ $R(\delta,(\mu_{1},\mu_{2}))$ becomes%
\begin{align}
R(\delta,(0,\mu_{2}))  &  =\mu_{2}\sum_{n_{2}=0}^{N_{2}-1}(1-n_{2}%
/N_{2})\binom{N_{2}}{n_{2}}\mu_{2}^{n_{2}}(1-\mu_{2})^{N_{2}-n_{2}}\nonumber\\
&  =\mu_{2}(1-\mu_{2})\sum_{n_{2}=0}^{N_{2}-1}\binom{N_{2}-1}{n_{2}}\mu
_{2}^{n_{2}}(1-\mu_{2})^{N_{2}-n_{2}-1}\nonumber\\
&  =\mu_{2}(1-\mu_{2})(\mu_{2}+(1-\mu_{2}))^{N_{2}-1}\nonumber\\
&  =\mu_{2}(1-\mu_{2}), \label{nature's best response}%
\end{align}
where in the second equation we use $(1-n_{2}/N_{2})\binom{N_{2}}{n_{2}%
}=\binom{N_{2}-1}{n_{2}}$ and the third equation uses the expansion of
$(\mu_{2}+(1-\mu_{2}))^{N_{2}-1}$ exploiting the binomial coefficients. Taking
FOC establishes that $\mu_{2}(1-\mu_{2})$ is maximized for $\mu_{2}=1/2$.

Next we show that $\delta$ is a best response to nature's strategy of mixing
uniformly over $(0,1/2)$ and $(1,1/2)$. Note that in both DGPs the value for
$\mu_{2}$ is the same. Because there are no observations for treatment 1 (and
the marginals $Y_{1}$ and $Y_{2}$ are independent), the sample conveys no
information about $\mu_{1}$. It follows that the conditional means for
outcomes under each treatment after observing any sample $w_{N}=(0,n_{2})$ are
the same as the unconditional ones and both equal $1/2$. Any treatment rule,
in particular the proposed rule, is therefore a best response. $\square
$\medskip

\textbf{Proof of Proposition \ref{symmetry statement}. (i) }Given the mixed
strategy $s\in\Delta\mathbb{S}$ by nature, iff a rule $\delta^{\circ}$ by the
policymaker satisfies the following condition
\begin{equation}
E(Y_{t}|w_{N})>E(Y_{t^{\prime}}|w_{N})\text{ then }\delta_{t^{\prime}}^{\circ
}(w_{N})=0 \label{optimality condition}%
\end{equation}
for any $t,t^{\prime}\in\{1,2\}$ and all samples $w_{N}$ with $P(w_{N})>0,$
then $\delta$ is a best response to $s$. We show below that for $s\in
\Delta\mathbb{S}$ as in (i) the following condition holds:%
\begin{equation}
E(Y_{1}|w_{N})-E(Y_{2}|w_{N})=E(Y_{2}|w_{N}^{\prime})-E(Y_{1}|w_{N}^{\prime})
\label{equality}%
\end{equation}
for any $w_{N}=(n_{1},n_{2})$ and $w_{N}^{\prime}=(n_{1}^{\prime}%
,n_{2}^{\prime})$ that satisfy (\ref{summability}). Assume $\delta^{\circ}$ is
a best response to $s.$ Then, by (\ref{optimality condition}), if
$E(Y_{t}|w_{N})>E(Y_{t^{\prime}}|w_{N})$ for a sample $w_{N}$ with
$P(w_{N})>0$ it must be that $\delta_{t^{\prime}}^{\circ}(w_{N})=0.$ But then,
by (\ref{equality}) $E(Y_{t^{\prime}}|w_{N}^{\prime})>E(Y_{t}|w_{N}^{\prime})$
and $\delta_{t^{\prime}}^{\circ}(w_{N}^{\prime})=1$ and therefore
(\ref{symmetry def}) holds. Lastly, if for a sample $w_{N}$ with $P(w_{N})>0,$
$E(Y_{t}|w_{N})=E(Y_{t^{\prime}}|w_{N})$ holds then, by (\ref{equality}) also
$E(Y_{t}|w_{N}^{\prime})=E(Y_{t^{\prime}}|w_{N}^{\prime})$ holds and
optimality of a treatment rule is not affected by what value the treatment
rule assigns to the samples $w_{N}$ and $w_{N}^{\prime},$ in particular, these
values can be picked to satisfy (\ref{symmetry def}).

We are therefore left to show (\ref{equality}). We first provide some
preliminary derivations. Note first that for $t\in\{1,2\}$ we have%
\begin{align}
&  E(Y_{t}|w_{N})\nonumber\\
&  \overset{}{=}P(Y_{t}\overset{}{=}1|w_{N})\nonumber\\
&  \overset{}{=}%
{\textstyle\sum\nolimits_{m=1}^{2\overline{m}}}
P(Y_{t}\overset{}{=}1\text{ \& nature picks }\overline{\mu}_{m}|w_{N}%
)\nonumber\\
&  \overset{}{=}%
{\textstyle\sum\nolimits_{m=1}^{2\overline{m}}}
P(Y_{t}\overset{}{=}1|\text{nature picks }\overline{\mu}_{m}\text{ \& }%
w_{N})P(\text{nature picks }\overline{\mu}_{m}|w_{N})\nonumber\\
&  \overset{}{=}%
{\textstyle\sum\nolimits_{m=1}^{2\overline{m}}}
P(Y_{t}\overset{}{=}1|\text{nature picks }\overline{\mu}_{m})P(w_{N}%
|\text{nature picks }\overline{\mu}_{m})p_{m}/P(w_{N})\nonumber\\
&  \overset{}{=}%
{\textstyle\sum\nolimits_{m=1}^{2\overline{m}}}
\mu_{tm}P(w_{N}|\text{nature picks }\overline{\mu}_{m})p_{m}/P(w_{N}),
\label{string of equ}%
\end{align}
where the first equality holds because $Y_{t}$ is a Bernoulli random variable,
the second equality holds by the law of total probability, the third one uses
the definition of conditional probability, the fourth one holds by Bayes'
theorem and exploits $P(Y_{t}=1|$nature picks $\overline{\mu}_{m}$ \&
$w_{N})=P(Y_{t}=1|$nature picks $\overline{\mu}_{m})$, $P(w_{N})>0,$ and
$P($nature picks $\overline{\mu}_{m})=p_{m}$ for all $m,$ and the fifth one
uses $P(Y_{t}=1|$nature picks $\overline{\mu}_{m})$ $=\mu_{tm}$.

Next, note that for $w_{N}=(n_{1},n_{2})$ and $w_{N}^{\prime}=(n_{1}^{\prime
},n_{2}^{\prime})$ that satisfy (\ref{summability}) we have for all
$m\leq\overline{m}$
\begin{align}
&  P(w_{N}|\text{ nature picks }\overline{\mu}_{m})\nonumber\\
&  =%
{\textstyle\prod\nolimits_{t=1}^{2}}
\binom{N_{t}}{n_{t}}\mu_{tm}^{n_{t}}(1-\mu_{tm})^{N_{t}-n_{t}}\nonumber\\
&  =%
{\textstyle\prod\nolimits_{t=1}^{2}}
\binom{N_{t}}{n_{t}}(1-\mu_{tm+\overline{m}})^{N_{t}-n_{t}^{\prime}}%
\mu_{tm+\overline{m}}{}^{n_{t}^{\prime}}\nonumber\\
&  =P(w_{N}^{\prime}|\text{nature picks }\overline{\mu}_{m+\overline{m}}),
\label{equ of cond prob}%
\end{align}
where the second equality uses (\ref{summability}) and (\ref{cond sym}).
Likewise, we have
\begin{equation}
P(w_{N}|\text{nature picks }\overline{\mu}_{m+\overline{m}})=P(w_{N}^{\prime
}|\text{nature picks }\overline{\mu}_{m}). \label{equ of cond prob 2}%
\end{equation}
Note that then by (\ref{cond sym})%
\begin{align}
&  P(w_{N})\nonumber\\
&  =%
{\textstyle\sum\nolimits_{m=1}^{\overline{m}}}
P(w_{N}|\text{nature picks }\overline{\mu}_{m})p_{m}+%
{\textstyle\sum\nolimits_{m=1}^{\overline{m}}}
P(w_{N}|\text{nature picks }\overline{\mu}_{m+\overline{m}})p_{m+\overline{m}%
}\nonumber\\
&  =%
{\textstyle\sum\nolimits_{m=1}^{\overline{m}}}
P(w_{N}^{\prime}|\text{nature picks }\overline{\mu}_{m+\overline{m}%
})p_{m+\overline{m}}+%
{\textstyle\sum\nolimits_{m=1}^{\overline{m}}}
P(w_{N}^{\prime}|\text{nature picks }\overline{\mu}_{m})p_{m}\nonumber\\
&  =P(w_{N}^{\prime}). \label{probs equ}%
\end{align}
Combining (\ref{equ of cond prob}), (\ref{cond sym}), and (\ref{probs equ}) we
obtain for all $m\leq\overline{m}$%
\begin{align}
P(w_{N}|\text{nature picks }\overline{\mu}_{m})p_{m}/P(w_{N})  &
=P(w_{N}^{\prime}|\text{nature picks }\overline{\mu}_{m+\overline{m}%
})p_{m+\overline{m}}/P(w_{N}^{\prime}),\nonumber\\
P(w_{N}|\text{nature picks }\overline{\mu}_{m+\overline{m}})p_{m+\overline{m}%
}/P(w_{N})  &  =P(w_{N}^{\prime}|\text{nature picks }\overline{\mu}_{m}%
)p_{m}/P(w_{N}^{\prime}). \label{key ingre}%
\end{align}
With these preliminaries we can now easily establish (\ref{equality}).
Namely,
\begin{align}
&  E(Y_{1}|w_{N})-E(Y_{2}|w_{N})\nonumber\\
&  =%
{\textstyle\sum\nolimits_{m=1}^{2\overline{m}}}
(\mu_{1m}-\mu_{2m})P(w_{N}|\text{nature picks }\overline{\mu}_{m}%
)p_{m}/P(w_{N})\nonumber\\
&  =%
{\textstyle\sum\nolimits_{m=1}^{\overline{m}}}
(\mu_{1m}-\mu_{2m})P(w_{N}^{\prime}|\text{nature picks }\overline{\mu
}_{m+\overline{m}})p_{m+\overline{m}}/P(w_{N}^{\prime})+\nonumber\\
&
{\textstyle\sum\nolimits_{m=1}^{\overline{m}}}
(\mu_{1m+\overline{m}}-\mu_{2m+\overline{m}})P(w_{N}^{\prime}|\text{nature
picks }\overline{\mu}_{m})p_{m}/P(w_{N}^{\prime})\nonumber\\
&  =%
{\textstyle\sum\nolimits_{m=1}^{2\overline{m}}}
(\mu_{2m}-\mu_{1m})P(w_{N}^{\prime}|\text{nature picks }\overline{\mu}%
_{m})p_{m}/P(w_{N}^{\prime})\nonumber\\
&  =E(Y_{2}|w_{N}^{\prime})-E(Y_{1}|w_{N}^{\prime}), \label{last step}%
\end{align}
where the first and last equalities hold by (\ref{string of equ}), the second
equality holds by (\ref{key ingre}), and the third one by (\ref{cond sym}).

\textbf{(ii)} By $\delta_{2}(n_{1},n_{2})$ we denote the probability with
which $\delta$ chooses treatment 2 when $n_{t}$ successes are observed for
$t=1,2.$ Note that
\begin{align}
&  E_{(\mu_{1},\mu_{2})}(\delta_{2}(w_{N}))\nonumber\\
&  =\sum_{n_{1}=0}^{N_{1}}\sum_{n_{2}=0}^{N_{2}}\delta_{2}(n_{1},n_{2}%
)\binom{N_{1}}{n_{1}}\mu_{1}^{n_{1}}(1-\mu_{1})^{N_{1}-n_{1}}\binom{N_{2}%
}{n_{2}}\mu_{2}^{n_{2}}(1-\mu_{2})^{N_{2}-n_{2}}\nonumber\\
&  =\sum_{k=0}^{N_{1}}\sum_{l=0}^{N_{2}}\delta_{2}(N_{1}-k,N_{2}%
-l)\binom{N_{1}}{N_{1}-k}\mu_{1}^{N_{1}-k}(1-\mu_{1})^{k}\binom{N_{2}}%
{N_{2}-l}\mu_{2}^{N_{2}-l}(1-\mu_{2})^{l}\nonumber\\
&  =\sum_{k=0}^{N_{1}}\sum_{l=0}^{N_{2}}(1-\delta_{2}(k,l))\binom{N_{1}}%
{N_{1}-k}\mu_{1}^{N_{1}-k}(1-\mu_{1})^{k}\binom{N_{2}}{N_{2}-l}\mu_{2}%
^{N_{2}-l}(1-\mu_{2})^{l}\nonumber\\
&  =1-E_{(1-\mu_{1},1-\mu_{2})}(\delta_{2}(w_{N})), \label{expe expre}%
\end{align}
where the second equation uses a change of the summation indices from
$N_{1}-n_{1}$ to $k$ and $N_{2}-n_{2}$ to $l$ and the third equality uses the
symmetry property of $\delta$ from (\ref{symmetry def}) and a well-known
property of the binomial coefficient. Note that%
\begin{align}
&  R(\delta,(\mu_{1},\mu_{2}))\nonumber\\
&  =\max\{\mu_{1},\mu_{2}\}-\mu_{1}E_{(\mu_{1},\mu_{2})}(\delta_{1}%
(w_{N}))-\mu_{2}E_{(\mu_{1},\mu_{2})}(\delta_{2}(w_{N}))\nonumber\\
&  =\max\{\mu_{1},\mu_{2}\}-\mu_{1}(1-E_{(\mu_{1},\mu_{2})}(\delta_{2}%
(w_{N})))-\mu_{2}E_{(\mu_{1},\mu_{2})}(\delta_{2}(w_{N})).
\label{formula regret}%
\end{align}
Wlog assume that $\mu_{1}<\mu_{2}$, in which case we obtain $R(\delta,(\mu
_{1},\mu_{2}))=(\mu_{2}-\mu_{1})(1-E_{(\mu_{1},\mu_{2})}(\delta_{2}(w_{N})))$.
Furthermore,%
\begin{align}
&  R(\delta,(1-\mu_{1},1-\mu_{2}))\nonumber\\
&  =\max\{1-\mu_{1},1-\mu_{2}\}-(1-\mu_{1})(1-E_{(1-\mu_{1},1-\mu_{2})}%
(\delta_{2}(w_{N})))-(1-\mu_{2})E_{(1-\mu_{1},1-\mu_{2})}(\delta_{2}%
(w_{N}))\nonumber\\
&  =(1-\mu_{1})-(1-\mu_{1})E_{(\mu_{1},\mu_{2})}(\delta_{2}(w_{N}%
)))-(1-\mu_{2})(1-E_{(\mu_{1},\mu_{2})}(\delta_{2}(w_{N})))\nonumber\\
&  =(\mu_{2}-\mu_{1})(1-E_{(\mu_{1},\mu_{2})}(\delta_{2}(w_{N}))),
\label{last step2}%
\end{align}
where the second equality uses (\ref{expe expre}). It follows that
$R(\delta,(1-\mu_{1},1-\mu_{2}))=R(\delta,(\mu_{1},\mu_{2})).$It follows that
when $(\mu_{1},\mu_{2})$ maximizes regret then so does $(1-\mu_{1},1-\mu_{2})$
if $\delta$ is chosen to satisfy (\ref{symmetry def}). $\square\medskip$

\textbf{Proof of Lemma \ref{lem:continuity in lead example}.} In general,%
\begin{align}
&  R(\delta,(\mu_{1},\mu_{2}))\nonumber\\
&  =\max\{\mu_{1},\mu_{2}\}-\mu_{1}E_{(\mu_{1},\mu_{2})}(\delta_{1}%
(w_{N}))-\mu_{2}E_{(\mu_{1},\mu_{2})}(\delta_{2}(w_{N}))
\label{general regret formula}%
\end{align}
and, under the particular sampling design considered here,
\begin{align}
&  E_{(\mu_{1},\mu_{2})}(\delta_{2}(w_{N}))\nonumber\\
&  =\sum_{n_{1}=0}^{N_{1}}\sum_{n_{2}=0}^{N_{2}}\delta_{2}(n_{1},n_{2}%
)\binom{N_{1}}{n_{1}}\mu_{1}^{n_{1}}(1-\mu_{1})^{N_{1}-n_{1}}\binom{N_{2}%
}{n_{2}}\mu_{2}^{n_{2}}(1-\mu_{2})^{N_{2}-n_{2}}. \label{form exp}%
\end{align}
Given that $\delta_{2}(n_{1},n_{2})\in\lbrack0,1]$ uniform continuity clearly
holds. $\square\medskip$

\textbf{Proof of Proposition \ref{symmetry statement in}. (i)} We will show
below that for $s\in\Delta\mathbb{S}$ as in Proposition
\ref{symmetry statement in}(i),
\begin{equation}
E(Y_{t}|w_{N})=E(Y_{\sigma^{-1}(t)}|w_{N}^{\prime})\text{ for }t\in
\{1,...,T-1\} \label{equality in}%
\end{equation}
holds for all samples $w_{N}^{\prime}$ as in (\ref{symmetry def in}) and
permutations $\sigma$ on $\{1,...,T-1\}$, where $\sigma^{-1}$ is the inverse
of permutation $\sigma$ defined on the group of permutations.

We first establish that (\ref{equality in}) implies that a symmetric best
response $\delta$ to nature's mixed strategy exists. Recall a treatment rule
$\delta^{\circ}$ is a best response to nature's strategy iff for all samples
$w_{N}$ with $P(w_{N})>0$
\begin{equation}
E(Y_{t}|w_{N})>E(Y_{t^{\prime}}|w_{N})\text{ implies }\delta_{t^{\prime}%
}^{\circ}(w_{N})=0 \label{optimality}%
\end{equation}
for $t,t^{\prime}\in\{1,...,T\}$. Note that $E(Y_{T}|w_{N})=\mu_{T}$ because
the mean of the status quo treatment is known.

Without loss of generality, assume that under sample $w_{N}$ we can order the
posterior mean as (with possible relabeling)
\begin{equation}
E(Y_{1}|w_{N})=\cdots=E(Y_{\bar{t}}|w_{N})>E(Y_{\bar{t}+1}|w_{N})\geq
\cdots\geq E(Y_{T-1}|w_{N}) \label{string of inequalities}%
\end{equation}
for some $\bar{t}<T-1$. Otherwise, the decision maker will be indifferent with
all $T-1$ treatments under both $w_{N}$ and $w_{N}^{\prime}$ in view of
(\ref{equality in}), so he can choose a treatment rule that satisfies the
symmetry condition (\ref{symmetry def in}).

Let $E^{max}$ be the value of $E(Y_{1}|w_{N})$. Using (\ref{equality in}), we
have
\begin{equation}
E^{max}=E(Y_{\sigma^{-1}(1)}|w_{N}^{\prime})=\cdots=E(Y_{\sigma^{-1}(\bar{t}%
)}|w_{N}^{\prime})>E(Y_{\sigma^{-1}(\bar{t}+1})|w_{N}^{\prime})\geq\cdots\geq
E(Y_{\sigma^{-1}(T-1)}|w_{N}^{\prime}) \label{ranking}%
\end{equation}

If $\mu_{T}>E^{max}$, then under both $w_{N}$ and $w_{N}^{\prime}$, define
$\delta$ to assign zero probability to the treatments $1,\cdots,T-1$ and
probability 1 to treatment $T.$ Then $\delta$ satisfies the symmetry restriction.

If $\mu_{T}<E^{max}$, then the policymaker is indifferent between treatments
$1,\cdots,\bar{t}$ under $w_{N}$ and indifferent between $\sigma
^{-1}(1),\cdots,\sigma^{-1}(\bar{t})$ under $w_{N}^{\prime}$, we can choose
$\delta$ as a treatment rule that respects the symmetry condition and puts
zero probability on treatments $\bar{t}+1,\cdots,T-1$ under sample $w_{N}$ and
zero probability on treatments $\sigma^{-1}(\bar{t}+1),\cdots,\sigma
^{-1}(T-1)$ under sample $w_{N}^{\prime}$.

If $\mu_{T}=E^{max}$, then the policy maker is indifferent between treatments
$1,\cdots,\bar{t},T$ under $w_{N}$ and indifferent between $\sigma
^{-1}(1),\cdots,\sigma^{-1}(\bar{t}),T$ under $w_{N}^{\prime}$. Using the same
reasoning as above, one can choose $\delta$ such that $\delta_{T}%
(w_{N})=\delta_{T}(w_{N}^{\prime})$ and the remaining treatment rule satisfies
the symmetry restrictions in (\ref{symmetry def in}).\smallskip

We now verify (\ref{equality in}). Take $w_{N}$ and $w_{N}^{\prime}$ such that
$n_{\sigma(t)}=n_{t}^{\prime}$\emph{ }for some permutation $\sigma$ on
$\{1,...,T-1\}$ as above (\ref{symmetry def in}). We will hold $\sigma$ fixed
for the rest of the proof. Define the shorthand notation $\mu_{m}=(\mu
_{1m},...,\mu_{T-1m},\mu_{T})$, $\mu_{\sigma m}=(\mu_{\sigma(1)m}%
,...,\mu_{\sigma(T-1)m},\mu_{T}),$ and
\begin{equation}
B(n,\mu_{tm})=\binom{\overline{N}}{n}\mu_{tm}^{n}(1-\mu_{tm})^{\overline{N}%
-n}. \label{Def B}%
\end{equation}
We have for any permutation $\sigma^{\prime}$
\begin{equation}
P(w_{N}|\text{nature picks }\mu_{ \sigma^{\prime}m})=%
{\textstyle\prod\nolimits_{t=1}^{T-1}}
B(n_{t},\mu_{\sigma^{\prime}m})=%
{\textstyle\prod\nolimits_{t=1}^{T-1}}
B(n_{\sigma(t)},\mu_{\sigma\sigma^{\prime}m})=P(w_{N}^{\prime}|\text{nature
picks }\mu_{\sigma\sigma^{\prime}m}), \label{Prob exp}%
\end{equation}
where the second equality uses a change in the order of multiplication, and
the last equality uses the definition of $w_{N}^{\prime}$.

Also,
\begin{align}
&  P(w_{N})\nonumber\\
&  =%
{\textstyle\sum\nolimits_{m=1}^{\overline{m}}}
p_{m}%
{\textstyle\sum\nolimits_{\sigma^{\prime}}}
((T-1)!)^{-1}P(w_{N}|\text{nature picks }\mu_{\sigma^{\prime}m})\nonumber\\
&  =%
{\textstyle\sum\nolimits_{m=1}^{\overline{m}}}
p_{m}%
{\textstyle\sum\nolimits_{\sigma^{\prime}}}
((T-1)!)^{-1}P(w_{N}^{\prime}|\text{nature picks }\mu_{\sigma\sigma^{\prime}%
m})\nonumber\\
&  =P(w_{N}^{\prime}), \label{prob equal}%
\end{align}
where the summation is over all permutations $\sigma^{\prime}$ on
$\{1,...,T-1\}$ and where we use (\ref{Prob exp}) and the fact that
$\sigma\sigma^{\prime}$ loops over all possible permutations as we vary
$\sigma^{\prime}$ in the second equality. The set of all permutations form a
group (with the obvious operation), so for each permutation $\sigma$, there
exists a unique inverse $\sigma^{-1}$. We obtain%
\begin{align}
&  E(Y_{t}|w_{N})\nonumber\\
&  =%
{\textstyle\sum\nolimits_{m=1}^{\overline{m}}}
{\textstyle\sum\nolimits_{\sigma^{\prime}}}
\mu_{\sigma^{\prime}(t)m}P(w_{N}|\text{nature picks }\mu_{\sigma^{\prime}%
m})p_{m}/((T-1)!P(w_{N}))\nonumber\\
&  =%
{\textstyle\sum\nolimits_{m=1}^{\overline{m}}}
[%
{\textstyle\sum\nolimits_{\sigma^{\prime}}}
\mu_{\sigma^{\prime}(t)m}P(w_{N}^{\prime}|\text{nature picks }\mu
_{\sigma\sigma^{\prime}m})]p_{m}/((T-1)!P(w_{N}^{\prime}))\nonumber\\
&  =%
{\textstyle\sum\nolimits_{m=1}^{\overline{m}}}
[%
{\textstyle\sum\nolimits_{\sigma^{\prime}}}
\mu_{\sigma^{-1}\sigma\sigma^{\prime}(t)m}P(w_{N}^{\prime}|\text{nature picks
}\mu_{\sigma\sigma^{\prime}m})]p_{m}/((T-1)!P(w_{N}^{\prime}))\nonumber\\
&  ={\sum\nolimits_{m=1}^{\overline{m}}}{\sum\nolimits_{\sigma\sigma^{\prime}%
}}\mu_{\sigma^{-1}\sigma\sigma^{\prime}(t)m}P(w_{N}^{\prime}|\text{nature
picks }\mu_{\sigma\sigma^{\prime}m})]p_{m}/((T-1)!P(w_{N}^{\prime
}))\nonumber\\
&  =E(Y_{\sigma^{-1}(t)}|w_{N}^{\prime}),
\end{align}
where the second equality uses (\ref{Prob exp}) and (\ref{prob equal}) and the
third equality uses the identity permutation can be written as multiplication
of $\sigma$ and $\sigma^{-1}$. The fourth equality uses $\sigma\sigma^{\prime
}$ loops over all permutations as we vary $\sigma^{\prime}$ and hold fixed
$\sigma$.

\textbf{(ii).} Let $\delta_{t}(n_{1},...,n_{T-1})$ be the probability that
treatment $t$ is chosen when $w_{N}=(n_{1},...,n_{T-1})$ is observed. Let
$\sigma$ be a permutation on $\{1,...,T-1\}.$ We use the shorthand notation
$\mu=(\mu_{1},...,\mu_{T-1},\mu_{T})$ and $\mu_{\sigma}=(\mu_{\sigma
(1)},...,\mu_{\sigma(T-1)},\mu_{T}).$ Note that for $t=1,...,T-1$%
\begin{align}
&  E_{\mu}(\delta_{t}(w_{N}))\nonumber\\
&  =\sum_{n_{1}=0}^{\overline{N}}...\sum_{n_{T-1}=0}^{\overline{N}}\delta
_{t}(n_{1},...,n_{T-1})B(n_{1},\mu_{1})...B(n_{T-1},\mu_{T-1})\nonumber\\
&  =\sum_{n_{1}=0}^{\overline{N}}...\sum_{n_{T-1}=0}^{\overline{N}}%
\delta_{\sigma^{-1}(t)}(n_{\sigma(1)},...,n_{\sigma(T-1)})B(n_{1},\mu
_{1})...B(n_{T-1},\mu_{T-1})\nonumber\\
&  =\sum_{n_{\sigma(1)}=0}^{\overline{N}}...\sum_{n_{\sigma(T-1)}%
=0}^{\overline{N}}\delta_{\sigma^{-1}(t)}(n_{\sigma(1)},...,n_{\sigma
(T-1)})B(n_{\sigma(1)},\mu_{\sigma(1)})...B(n_{\sigma(T-1)},\mu_{\sigma
(T-1)})\nonumber\\
&  =\sum_{n_{1}=0}^{\overline{N}}...\sum_{n_{T-1}=0}^{\overline{N}}%
\delta_{\sigma^{-1}(t)}(n_{1},...,n_{T-1})B(n_{1},\mu_{\sigma(1)}%
)...B(n_{T-1},\mu_{\sigma(T-1)})\nonumber\\
&  =E_{\mu_{\sigma}}(\delta_{\sigma^{-1}(t)}(w_{N})), \label{expe expre in}%
\end{align}
where (\ref{symmetry def in}) is used in the second equality, the third
equality simply changes the order of summation, and the fourth equality
follows from the change in variables $n_{\sigma(t)}\rightarrow n_{t}$ for
$t=1,...,T-1$. The previous derivation implies that also $E_{\mu}(\delta
_{T}(w_{N}))=E_{\mu_{\sigma}}(\delta_{T}(w_{N})).$ Note that%
\begin{align}
&  R(\delta,\mu_{\sigma})\nonumber\\
&  =\max\{\mu_{\sigma(1)},...,\mu_{\sigma(T-1)},\mu_{T}\}-%
{\textstyle\sum\nolimits_{t=1}^{T-1}}
\mu_{\sigma(t)}E_{\mu_{\sigma}}(\delta_{t}(w_{N}))-\mu_{T}E_{\mu_{\sigma}%
}(\delta_{T}(w_{N}))\nonumber\\
&  =\max\{\mu_{\sigma(1)},...,\mu_{\sigma(T-1)},\mu_{T}\}-{\sum\nolimits_{t=1}%
^{T-1}}\mu_{\sigma(t)}E_{\mu_{\sigma}}(\delta_{\sigma^{-1}\sigma(t)}%
(w_{N}))-\mu_{T}E_{\mu_{\sigma}}(\delta_{T}(w_{N}))\nonumber\\
&  =\max\{\mu_{1},...,\mu_{T}\}-%
{\textstyle\sum\nolimits_{t=1}^{T-1}}
\mu_{\sigma(t)}E_{\mu}(\delta_{\sigma(t)}(w_{N}))-\mu_{T}E_{\mu}(\delta
_{T}(w_{N}))\nonumber\\
&  =\max\{\mu_{1},...,\mu_{T}\}-%
{\textstyle\sum\nolimits_{t=1}^{T-1}}
\mu_{t}E_{\mu}(\delta_{t}(w_{N}))-\mu_{T}E_{\mu}(\delta_{T}(w_{N}))\nonumber\\
&  =R(\delta,\mu), \label{regret equal}%
\end{align}
where we used (\ref{expe expre in}) in the third equality and a change in the
order of summation in the fourth equality. Consequently, if $\mu$ maximizes
regret given the symmetric $\delta$, then $\mu_{\sigma}$ also maximizes the
regret for $\delta$. $\square\medskip$

\textbf{Proof of Lemma \ref{lem:continuity in innovation}.} Note that
\begin{equation}
R(\delta,\mu)=\max_{t=1,...,T}\{\mu_{t}\}-%
{\textstyle\sum\nolimits_{t=1}^{T-1}}
\mu_{t}E_{\mu}(\delta_{t}(w_{N}))-\mu_{T}(1-%
{\textstyle\sum\nolimits_{t=1}^{T-1}}
E_{\mu}(\delta_{t}(w_{N}))) \label{regret formula in}%
\end{equation}
and, under the particular sampling design considered here, for $t\in
\{1,...,T-1\}$
\begin{equation}
E_{\mu}(\delta_{t}(w_{N}))=\sum_{n_{1}=0}^{\overline{N}}...\sum_{n_{T-1}%
=0}^{\overline{N}}\delta_{t}(n_{1},...,n_{T-1})B(n_{1},\mu_{1})...B(n_{T-1}%
,\mu_{T-1}). \label{form exp in}%
\end{equation}
Given that $\delta_{t}(n_{1},...,n_{T-1})\in\lbrack0,1]$ uniform continuity
clearly holds. $\square$\medskip\bigskip

{\Large References}

\begin{description}
\item Aradillas Fern\'{a}ndez, A., J. Blanchet, J.L. Montiel Olea, C. Qiu, J.
Stoye, and L. Tan (2024), \textquotedblleft$\epsilon$-Minimax Solutions of
Statistical Decision Problems,\textquotedblright\ unpublished working paper,
Cornell University.

\item Aradillas Fern\'{a}ndez, A., J.L. Montiel Olea, C. Qiu, J. Stoye, and S.
Tinda (2024), \textquotedblleft Robust Bayes Treatment Choice with Partial
Identification,\textquotedblright\ arXiv:2408.11621.

\item Ben-Tal, A., T. Margalit, and A. Nemirovski (2001), \textquotedblleft
The ordered subsets mirror descent optimization method with applications to
tomography,\textquotedblright\ \emph{SIAM Journal on Optimization}, 12, 79--108.

\item Berger, J. (1985), Statistical Decision Theory and Bayesian Analysis,
Second Edition, New York: SpringerVerlag.

\item Bubeck, S. (2015), Convex Optimization: Algorithms and Complexity.
Foundations and Trends in Machine Learning, vol. 8, no. 3-4, 231--358.

\item Chamberlain, G. (2000), \textquotedblleft Econometric Applications of
Maxmin Expected Utility,\textquotedblright\ \emph{Journal of Applied
Econometrics}, 15, 625--644.

\item Chen, H. and P. Guggenberger (2024), \textquotedblleft A note on minimax
regret rules with multiple treatments in finite samples\textquotedblright,
forthcoming in \emph{Econometric Theory}.

\item Cucconi, O. (1968), \textquotedblleft Contributi all'Analisi Sequenziale
nel Controllo di Accettazione per Variabili. \emph{Atti dell' Associazione
Italiana per il Controllo della Qualit\`{a}} 6, 171--186.

\item Daskalakis, C. and Q. Pan (2014), \textquotedblleft A Counter-Example to
Karlin's Strong Conjecture for Fictitious Play,\textquotedblright\ 55th IEEE
Symposium on Foundations of Computer Science, FOCS.

\item Daskalakis, C., A. Deckelbaum, and A. Kim (2024), \textquotedblleft
Near-Optimal No-Regret Algorithms for Zero-Sum Games,\textquotedblright%
\ unpublished working paper, MIT.

\item Dominitz, J. and C. Manski (2024), \textquotedblleft Comprehensive OOS
Evaluation of Predictive Algorithms with Statistical Decision
Theory\textquotedblright, unpublished working paper, Northwestern University.

\item Elliott, G., U. M\"{u}ller, and M.W.Watson (2015), \textquotedblleft
Nearly Optimal Tests when a Nuisance Parameter is Present Under the Null
Hypothesis,\textquotedblright\ \emph{Econometrica}, 83, 771--811.

\item Fudenberg, D. and D.K. Levine (1998). The theory of learning in games.
Volume 2. MIT press.

\item Giacomini, R. and T. Kitagawa (2022), \textquotedblleft Robust Bayesian
inference for set-identified models\textquotedblright, \emph{Econometrica} 89, 1519--1556.

\item Guggenberger, P., N. Mehta, and N. Pavlov (2024), \textquotedblleft
Minimax regret treatment rules with finite samples when a quantile is the
object of interest,\textquotedblright\ unpublished working paper, Pennsylvania
State University.

\item Gupta, S. and S. Hande (1992), \textquotedblleft On some nonparametric
selection procedures. In: Saleh, A.K.Md.E. (Ed.), Nonparametric Statistics and
Related Topics. Elsevier.

\item Hirano, K. and J. Porter (2009), \textquotedblleft Asymptotics for
Statistical Treatment Rules,\textquotedblright\ \emph{Econometrica} 77, 1683--1701.

\item Ishihara, T. and T. Kitagawa (2021), \textquotedblleft Evidence
Aggregation for Treatment Choice,\textquotedblright\ unpublished working
paper, Brown University. ArXiv: 2108.06473.

\item Kempthorne, P.J. (1987), \textquotedblleft Numerical specification of
discrete least favorable prior distributions,\textquotedblright\ \emph{SIAM
Journal on Scientific and Statistical Computing}, 8, 171--184.

\item Kitagawa, T. and A. Tetenov (2018), \textquotedblleft Who Should be
Treated? Empirical Welfare Maximization Methods for Treatment
Choice,\textquotedblright\ \emph{Econometrica}, 86, 591--616.

\item Kitagawa, T., S. Lee, and C. Qiu (2024), \textquotedblleft Treatment
Choice with Nonlinear Regret,\textquotedblright\ arXiv:2205.08586.

\item Leslie, D.S and E.J. Collins (2006), \textquotedblleft Generalised
weakened fictitious play,\textquotedblright\ \emph{Games and Economic
Behaviour}, 56, 285--298.

\item Manski, C. (1988), \textquotedblleft Ordinal Utility Models of Decision
Making Under Uncertainty,\textquotedblright\ \emph{Theory and Decision}, 25, 79--104.

\item \_\_\_\_\_\_ (2004), \textquotedblleft Statistical Treatment Rules for
Heterogeneous Populations,\textquotedblright\ \emph{Econometrica}, 72, 221--246.

\item \_\_\_\_\_\_(2007), \textquotedblleft Minimax-regret treatment choice
with missing outcome data,\textquotedblright\ \emph{Journal of Econometrics,}
139, 105--115.

\item \_\_\_\_\_\_(2021), \textquotedblleft Econometrics for decision making:
Building foundations sketched by Haavelmo and Wald,\textquotedblright%
\ \emph{Econometrica,} 89, 2827--2853.

\item \_\_\_\_\_\_ and A. Tetenov (2016), \textquotedblleft Sufficient Trial
Size to Inform Clinical Practice,\textquotedblright\ \emph{Proceedings of the
National Academy of Sciences,} 113, 10518--10523.

\item \_\_\_\_\_\_\_\_\_\_\_\_(2019), \textquotedblleft Trial Size for
Near-Optimal Choice Between Surveillance and Aggressive Treatment:
Reconsidering MSLT-II,\textquotedblright\ \emph{The American Statistician} 73, 305--311.

\item \_\_\_\_\_\_\_\_\_\_\_\_(2021), \textquotedblleft Statistical Decision
Properties of Imprecise Trials Assessing Coronavirus Disease 2019 (COVID-19)
Drugs, \emph{Value in Health,} 24, 641--647.

\item \_\_\_\_\_\_\_\_\_\_\_\_(2023), \textquotedblleft Statistical Decision
Theory Respecting Stochastic Dominance,\textquotedblright\ \emph{The Japanese
Economic Review}, 74, 447--469.

\item Masten, M. (2023), \textquotedblleft Minimax-regret treatment rules with
many treatments,\textquotedblright\ \emph{The Japanese Economic Review}, 74, 501--537.

\item Montiel Olea, J.L, C. Qiu, and J. Stoye (2023), \textquotedblleft
Decision Theory for Treatment Choice with Partial
Identification,\textquotedblright\ ArXiv: 2312.17623.

\item Robinson, J. (1951), \textquotedblleft An iterative method of solving a
game,\textquotedblright\ \emph{Annals of Mathematics}, 54(2):296--301.

\item Savage, L. (1954), The Foundations of Statistics, New York: Wiley.

\item Schlag, K. (2003), \textquotedblleft How to minimize maximum regret
under repeated decision-making,\textquotedblright\ EUI working paper.

\item \_\_\_\_\_\_ (2006), \textquotedblleft ELEVEN - Tests needed for a
recommendation,\textquotedblright\ EUI working paper, ECO No. 2006/2.

\item Stoye, J. (2009), \textquotedblleft Minimax Regret Treatment Choice with
Finite Samples,\textquotedblright\ \emph{Journal of Econometrics}, 151, 70--81.

\item \_\_\_\_\_\_ (2012), \textquotedblleft Minimax Regret Treatment Choice
with Covariates or with Limited Validity of Experiments,\textquotedblright%
\ \emph{Journal of Econometrics}, 166, 138--156.

\item Tetenov, A. (2012), \textquotedblleft Statistical Treatment Choice Based
on Asymmetric Minimax Regret Criteria,\textquotedblright\ \emph{Journal of
Econometrics}, 166, 157--165.

\item Wald, A. (1950), Statistical Decision Functions, New York: Wiley.

\item Yata, Kohei (2023), \textquotedblleft Optimal Decision Rules Under
Partial Identification,\textquotedblright\ ArXiv:2111.04926.\pagebreak
\end{description}

\section{Supplementary Appendix}

The Supplementary Appendix provides some additional results not reported in
the main body of the paper. Namely, we report additional results from the
section on treatment assignment with unbalanced samples and all results from
the section on testing the status quo against two innovations.

\subsection{Additional Results for Treatment assignment with unbalanced
samples}

In this subsection we report i) results for additional sample sizes
$(N_{1},N_{2})$ not covered in the main body of the paper in TABLE\ I, ii)
results for additional sample sizes $(N_{1},N_{2})$ not covered in the main
body of the paper in TABLE\ II, and iii) an example of $\delta^{n}$.

i) \textbf{TABLE\ I (continued):} Maximal regret of $\delta^{n}$ for different
sample sizes, number of iterations, weighting schemes and initializations.%

\begin{tabular}
[c]{c||cccc}%
\ \ Weighting choice $\alpha_{n}$ \  & $\ \ \ \ \ \ n^{-1}$ \ \ \ \ \ \  &
$\ \ \ \ \ \ n^{-1}$ \ \ \ \  & $(5+n)^{-.7}$ \  & $\ \ (5+n)^{-.7}$\\
Initialization $\delta^{1}$ & SO & ES & SO & ES\\\hline\hline
\end{tabular}

$(N_{1},N_{2})=80;$\ minimax regret value=.01344651%

\begin{tabular}
[c]{c||cccc}\hline\hline
Maximal regret of $\delta^{1}$ & .3949134 & \textbf{.01}46335 & .3949134 &
\textbf{.01}46335\\
Maximal regret of $\delta^{150}$ & \textbf{.01}45548 & \textbf{.01}57714 &
\textbf{.0134}618 & \textbf{.0134}659\\
Maximal regret of $\delta^{500}$ & \textbf{.013}7163 & \textbf{.013}9940 &
\textbf{.01344}73 & \textbf{.0134}778\\
Maximal regret of $\delta^{2000}$ & \textbf{.013}5119 & \textbf{.013}5638 &
\textbf{.013446}7 & \textbf{.0134}617\\\hline\hline
\end{tabular}

$(N_{1},N_{2})=40;$\ minimax regret value=.01902905%

\begin{tabular}
[c]{c||cccc}\hline\hline
Maximal regret of $\delta^{1}$ & .3712294 & \textbf{.0}212825 & .3712294 &
\textbf{.0}212825\\
Maximal regret of $\delta^{150}$ & \textbf{.019}5494 & \textbf{.0}203048 &
.\textbf{0190}365 & \textbf{.0190}399\\
Maximal regret of $\delta^{500}$ & \textbf{.019}1818 & \textbf{.019}3110 &
\textbf{.0190}334 & \textbf{.0190}980\\
Maximal regret of $\delta^{2000}$ & \textbf{.0190}664 & \textbf{.019}1008 &
\textbf{.019029}6 & \textbf{.019029}2\\\hline\hline
\end{tabular}

$(N_{1},N_{2})=20;$\ minimax regret value=.02694711%

\begin{tabular}
[c]{c||cccc}\hline\hline
Maximal regret of $\delta^{1}$ & .3493432 & \textbf{.0}311654 & .3493432 &
\textbf{.0}311654\\
Maximal regret of $\delta^{150}$ & \textbf{.02}75462 & \textbf{.02}77504 &
\textbf{.0269}735 & \textbf{.0269}522\\
Maximal regret of $\delta^{500}$ & \textbf{.02}71285 & \textbf{.02}71833 &
\textbf{.0269}663 & \textbf{.02694}82\\
Maximal regret of $\delta^{2000}$ & \textbf{.0269}900 & \textbf{.02}70054 &
\textbf{.0269471} & \textbf{.026947}7\\\hline\hline
\end{tabular}

$(N_{1},N_{2})=10;$\ minimax regret value=.03820907%

\begin{tabular}
[c]{c||cccc}\hline\hline
Maximal regret of $\delta^{1}$ & \textbf{.3}353402 & \textbf{.0}460039 &
.3353402 & .\textbf{0}460039\\
Maximal regret of $\delta^{150}$ & \textbf{.038}6786 & \textbf{.038}7004 &
\textbf{.0382}168 & \textbf{.0382}227\\
Maximal regret of $\delta^{500}$ & \textbf{.038}3319 & \textbf{.038}3552 &
\textbf{.0382}167 & \textbf{.038209}1\\
Maximal regret of $\delta^{2000}$ & \textbf{.0382}395 & \textbf{.0382}450 &
\textbf{.0382090} & \textbf{.0382090}\\\hline\hline
\end{tabular}
\bigskip

ii) \textbf{TABLE II (continued): }Maximal regret of $\delta^{n}$ and
$R(\delta^{n},\mu_{BR}^{n})-R^{n}$ for various choices of $n$ and
$(N_{1},N_{2})$%

\begin{tabular}
[c]{c||ccc}%
$n$%
$\backslash$%
$(N_{1},N_{2})$ & $(50,60)$ & $(50,100)$ & $(50,150)$\\\hline\hline
$1$ & .016376;.012255 & .015723;.014738 & .014585;.013844\\
$150$ & .016272;.001074 & .014721;.000518 & .013886;.000424\\
$500$ & .016250;.000557 & .014706;.000545 & .013894;.001413\\
$2000$ & .016241;.000030 & .014705;.000208 & .013850;.000174\\
$I_{2000}$ & [.016225,.016241] & [.014647,.014703] &
[.013781,.013849]\\\hline\hline
\end{tabular}

\bigskip

iii) Next, we report $\delta^{5000}$ for $N_{1}=10$ and $N_{2}=20$, ES,\ and
$\alpha_{n}=(5+n)^{-.7}$. Namely, we tabulate $\delta_{2}^{5000}(n_{1},n_{2})$
in a matrix with $N_{1}+1=11$ rows and $N_{2}+1=21$ columns. The
$(i,j)$-element in the following matrix represents $\delta_{2}^{5000}%
(i-1,j-1),$ $i=1,...,11$ and $j=1,...,21.$

.0013 1 1.000 1 1.000 1 1.000 1 1.000 1 1.0 1 1.000 1 1.000 1 1.000 1 1.000 1 1.000

.0006 0 .9998 1 1.000 1 1.000 1 1.000 1 1.0 1 1.000 1 1.000 1 1.000 1 1.000 1 1.000

.0000 0 .0000 0 .4291 1 1.000 1 1.000 1 1.0 1 1.000 1 1.000 1 1.000 1 1.000 1 1.000

.0000 0 .0000 0 .0000 0 .4111 1 1.000 1 1.0 1 1.000 1 1.000 1 1.000 1 1.000 1 1.000

.0000 0 .0000 0 .0000 0 .0000 0 .4112 1 1.0 1 1.000 1 1.000 1 1.000 1 1.000 1 1.000

.0000 0 .0000 0 .0000 0 .0000 0 .0000 0 .50 1 1.000 1 1.000 1 1.000 1 1.000 1 1.000

.0000 0 .0000 0 .0000 0 .0000 0 .0000 0 .00 0 .5889 1 1.000 1 1.000 1 1.000 1 1.000

.0000 0 .0000 0 .0000 0 .0000 0 .0000 0 .00 0 .0000 0 .5889 1 1.000 1 1.000 1 1.000

.0000 0 .0000 0 .0000 0 .0000 0 .0000 0 .00 0 .0000 0 .0000 0 .5709 1 1.000 1 1.000

.0000 0 .0000 0 .0000 0 .0000 0 .0000 0 .00 0 .0000 0 .0000 0 .0000 0 .0002 1 .9994

.0000 0 .0000 0 .0000 0 .0000 0 .0000 0 .00 0 .0000 0 .0000 0 .0000 0 .0000 0 .9987

\subsection{Additional results for treatment assignment when testing
innovations\label{results in}}

We implement Algorithm \ref{specialized algorithm in} for the case $T=3$,
sample sizes $\overline{N}=N_{1}=N_{2}\in\{5,10,20,30,$ $40,50,100,200\}$ and
known mean of the status quo treatment $\mu_{3}\in\{.2,.5,.8\}.$ The
initialization rule is the empirical success rule described in the beginning
of Algorithm \ref{specialized algorithm in}. We choose $p=1000$ and
$\alpha_{n}=(5+n)^{-.7}$. TABLE IV (continued) reports\textbf{ }maximal regret
of $\delta^{n}$ and $R(\delta^{n},\mu_{BR}^{n})-R^{n}$ for various choices of
$n$ and $I_{2000}$ for $\overline{N}=N_{1}=N_{2}\in\{5,20,30,40\}.$

Note that if $\mu_{T}=0$ then it is always (at least weakly) better for the
policymaker to switch to one of the alternative treatment. Therefore, in that
case the problem is reduced to choosing between $T-1$ treatments which (for
the case of equal sample sizes) has been dealt with in Chen and Guggenberger
(2024). On the other hand, if $\mu_{T}=1$ it is always (at least weakly)
better for the policymaker to remain with the status quo no matter what the
data says. Therefore, it is enough to consider mean values $0<\mu_{T}%
<1$.\medskip

\textbf{TABLE IV (continued): }Maximal regret of $\delta^{n}$ and
$R(\delta^{n},\mu_{BR}^{n})-R^{n}$ for several $n$ and $I_{2000}$ for several
$\overline{N}$ and $\mu_{3}.$%

\begin{tabular}
[c]{c||ccc}\hline\hline
$\quad\ \mu_{3}\qquad$ & \quad\ \ \ \ \ \ \ \ 0.2 \ \ \ \ \ \quad &
\quad\ \ \ \ \ \ \ \ \ 0.5 \ \ \ \ \ \ \quad & \quad\ \ \ \ \ \ \ \ \ 0.8
\ \ \ \ \ \ \quad
\end{tabular}

\begin{tabular}
[c]{c||ccc}\hline\hline
\multicolumn{4}{l}{$\overline{N}=40$}\\\hline\hline
$n=1$ & .020020;.004062 & .023973;.023973 & .021501;.021501\\
$n=150$ & .019029;.000012 & .023511;.000744 & .018796;.000620\\
$n=500$ & .019029;.000021 & .023491;.000058 & .018931;.001456\\
$n=2000$ & .019029;.000000 & .023443;.000161 & .018725;.000125\\
$I_{2000}$ & [.019028;.019029] & [.023434;.023438] & [.018673;.018707]
\end{tabular}

\begin{tabular}
[c]{c||ccc}\hline\hline
\multicolumn{4}{l}{$\overline{N}=30$}\\\hline\hline
$n=1$ & .023387;.004777 & .027684;.027684 & .025266;.025266\\
$n=150$ & .021991;.000051 & .027122;.000160 & .021773;.000514\\
$n=500$ & .021987;.000042 & .027169;.000359 & .021721;.001021\\
$n=2000$ & .021982;.000009 & .027075;.000223 & .021625;.000078\\
$I_{2000}$ & [.021979;.021982] & [.027060;.027061] & [.021593;.021603]
\end{tabular}

\begin{tabular}
[c]{c||ccc}\hline\hline
\multicolumn{4}{l}{$\overline{N}=20$}\\\hline\hline
$n=1$ & .029221;.006025 & .033911;.033911 & .031851;.031851\\
$n=150$ & .027047;.000061 & .033271;.001901 & .026575;.000348\\
$n=500$ & .027025;.000027 & .033155;.000083 & .026521;.000120\\
$n=2000$ & .027009;.000011 & .033185;.000080 & .026492;.000256\\
$I_{2000}$ & [.026999;.027000] & [.033133;.033143] & [.026470;.026476]
\end{tabular}

\begin{tabular}
[c]{c||ccc}\hline\hline
\multicolumn{4}{l}{$\overline{N}=5$}\\\hline\hline
$n=1$ & .065019;.014349 & .064435;.010440 & .073224;.073224\\
$n=150$ & .057589;.000655 & .064419;.000924 & .053662;.000179\\
$n=500$ & .057316;.000276 & .064210;.000140 & .053707;.000225\\
$n=2000$ & .057185;.000167 & .064235;.000047 & .053740;.000163\\
$I_{2000}$ & [.057086;.057086] & [.064190;.064190] & [.053593;.053593]
\end{tabular}

\end{document}